\documentclass[11pt]{article}

\usepackage{epsfig,dsfont}
\usepackage{geometry}
\usepackage{lscape}
\usepackage{afterpage}
\usepackage{amsmath}
\usepackage[square,comma,numbers,sort&compress]{natbib}
\usepackage{float}
\usepackage{amssymb}

\newcommand*{\no}{\noindent}
\newcommand*{\bea}{\begin{eqnarray}}
\newcommand*{\eea}{\end{eqnarray}}
\newcommand*{\be}{\begin{equation}}
\newcommand*{\ee}{\end{equation}}

\newcommand*{\pref}[1]{(\ref{#1})}

\newcommand*{\nn}{\nonumber}

\newcommand{\bma}{\begin{pmatrix}}
\newcommand{\ema}{\end{pmatrix}}

\begin{document}  
 
\vspace*{1mm}

\begin{center}

{\LARGE Constructing a neutron star from the lattice in G$_2$-QCD}
\vskip10mm
Ouraman Hajizadeh, Axel Maas
\vskip8mm
University of Graz, Institute of Physics, NAWI Graz, Universit\"atsplatz 5, 8010 Graz, Austria
\end{center}
\vskip15mm

\begin{abstract}
The inner structure of neutron stars is still an open question. One obstacle is the infamous sign problem of lattice QCD, which bars access to the high-density equation of state. A possibility to make progress and understand the qualitative impact of gauge interactions on the neutron star structure is to study a modified version of QCD without the sign problem. In the modification studied here the gauge group of QCD is replaced by the exceptional Lie group G$_2$, which keeps neutrons in the spectrum. Using an equation of state from lattice calculations only we determine the mass-radius-relation for a neutron star using the Tolman-Oppenheimer-Volkoff equation. This allows us to understand the challenges and approximations currently necessary to use lattice data for this purpose. We discuss in detail the particular uncertainties and systematic problems of this approach.
\end{abstract}

\vskip15mm

\section{Introduction}

The properties of neutron stars \cite{Glendenning:1997wn,Steiner:2010fz,Kapusta:2006pm} have been the subject of many investigations for decades, both in astrophysics and particle physics. In addition, a vast database on neutron stars is nowadays available from astronomical observations \cite {Lattimer:2006xb,Glendenning:1997wn,Steiner:2010fz} and the potential of investigations using gravitational waves \cite{Abbott:2016blz} is very exciting  \cite{DelPozzo:2013ala}.

Nevertheless, a quantitative description of their properties from first principles, i.\ e.\ QCD, has not yet been achieved. The main reason for this is that it was not yet possible to derive the equation of state governing neutron stars from QCD at finite density and small or zero temperature reliably \cite{Friman:2011zz}. The origin of this deficiency is that the mainstay of non-perturbative QCD calculations, lattice gauge theory, is hampered by the sign problem, effectively making even qualitative statements almost impossible yet \cite{Gattringer:2010zz,deForcrand:2010ys,Aarts:2017vrv}. Alternative methods not facing the sign problem are either still not sufficiently far developed or make heavy use of modeling, which reduces their predictivity \cite{Leupold:2011zz,Buballa:2003qv,Pawlowski:2010ht,Braun:2011pp}.

Turning this around, data of neutron stars could provide us with urgently needed understanding of strong interactions in this cold and dense regime of the phase diagram \cite {Lattimer:2006xb,Glendenning:1997wn,Steiner:2010fz}, even more so with the yet untapped potential from gravitational waves. Hence, they make it possible to observe large-scale effects of non-Abelian gauge theories.

Pending a solution for QCD, several questions arise, which may be accessed using QCD-like theories. Especially interesting are two of them:
\begin{itemize}
 \item What is necessary for an ab-initio calculation to provide a suitable good result to describe a neutron star?
 \item Are there any generic signatures of non-Abelian gauge theories which can manifest in the features of neutron stars?
\end{itemize}
To answer them requires QCD-like theories, which are easier to treat than QCD at finite density, while at the same time are not too different from QCD itself. The most important constraint in the selection of a suitable QCD-like theory is the existence of fermionic baryons, especially neutrons. Otherwise, there will be no hadronic Fermi surface possible, which could stabilize a neutron star against further collapse\footnote{Not withstanding the possibility of quark stars.}. In addition, all other features should be as similar to QCD as possible, while the theory should still be tractable.

A theory which rather well satisfies these constraints is G$_2$-QCD, i.\ e.\ QCD where the gauge group SU(3) is replaced by the exceptional Lie group G$_2$ \cite{Holland:2003jy,Maas:2012wr}, see \cite{Maas:2012ts} for a review of the properties of this theory. This theory has been studied on the lattice, both with \cite{Maas:2012wr,Wellegehausen:2013cya,Wellegehausen:2015iea} and without quarks \cite{Pepe:2006er,Greensite:2006sm,Maas:2007af,Cossu:2007dk,Danzer:2008bk,Liptak:2008gx,Maas:2010qw,Wellegehausen:2009rq,Wellegehausen:2010ai,Ilgenfritz:2012aa,Bruno:2014rxa,Bonati:2015uga} and even with scalars \cite{Wellegehausen:2011jz}. In particular, information on the phase diagram and thermodynamic quantities \cite{Maas:2012wr,Wellegehausen:2013cya} and the spectrum \cite{Wellegehausen:2013cya} is available for this theory from lattice simulations. For completeness and to make this presentation self-contained for a wider audience, we rehearse the most pertinent features of G$_2$-QCD in section \ref{s:g2qcd}.
 
Based on these results, we determine the mass-radius relation of a G$_2$-QCD neutron star, using the Tolman-Oppenheimer-Volkoff (TOV) equation. The details of our approach will be given in section \ref{s:mr}. In response to the first question, this requires a number of approximations, which will also be described there. This will show the particular problems induced by employing results of a lattice calculation. These go beyond the usual lattice-inherent systematic and statistical uncertainties, which add to the uncertainties as well. All these error sources require therefore study, and we will do so in detail in section \ref{s:sys}. Thus, the results will be even for G$_2$-QCD qualitative, rather than (semi-)quantitative.

However, in response to the second question, we observe in section \ref{s:ns} a number of qualitative features. Especially the interaction structure seems to be able to imprint itself, though washed-out, in the mass-radius relation. This is a very exciting hint, as the question whether neutron stars have a homogeneous structure or a layered structure is one of the central questions. These results indicate that this question may already be solvable using astronomical data only. Of course, such statements could also be obtained from analyzing a model equation of state. It can also be improved by supplementing results from other methods, especially nuclear physics \cite{Glendenning:1997wn,Steiner:2010fz,Kapusta:2006pm}. But aside from the yet unknown nuclear physics of G$_2$-QCD \cite{Wellegehausen:2013cya}, here the central aim is to see how far it is possible to go by using only lattice  results. And thus the result here is a statement based, for the first time, on a first-principle lattice evaluation of the equation-of-state, and thus a genuine non-Abelian gauge theory.

In total, the features we obtained are not too different from the expectations of real neutron stars \cite{Glendenning:1997wn,Steiner:2010fz,Kapusta:2006pm}, which by itself is promising. Given that these are consistent with the astrophysical data, this also implies that the G$_2$ neutron star is perhaps not that different from the QCD one. This would suggest that QCD-like theories may have more to say on the second question than expected.

The results will be summarized in section \ref{s:sum}. Note that some preliminary results can be found in \cite{Hajizadeh:2016jvj}.
 
\section{G$_2$-QCD and degrees of freedom}\label{s:g2qcd}

G$_2$-QCD is ordinary QCD, but with the gauge group SU(3) replaced by the exceptional Lie group G$_2$ \cite{Holland:2003jy,Maas:2012wr,Maas:2012ts}. This is the only change, and all other features are results of this change.

G$_2$ is a subgroup of SO(7). There are 7 quark colors and 14 gluons \cite{Holland:2003jy}. Note that all representations are real. Therefore it does not suffer from a sign problem on the lattice \cite{Maas:2012wr,Kogut:2000ek}. The disadvantage is that quarks and antiquarks are not independent, but related by charge conjugation, which has a number of implications \cite{Holland:2003jy,Wellegehausen:2013cya,Maas:2012ts}, similar to the two-color case \cite{Kogut:2000ek}. In particular, this implies an enlarged chiral symmetry SU($2N_f$) for $N_f$ flavors, the Pauli-G\"ursey symmetry. This makes a non-trivial breaking of this symmetry possible even for a single flavor \cite{Holland:2003jy,Maas:2012wr,Wellegehausen:2013cya,vonSmekal:2012vx}. Note that the Goldstone bosons in the one-flavor case are diquarks, while they are both diquarks and mesons for more than one flavor \cite{Wellegehausen:2013cya,Kogut:2000ek,vonSmekal:2012vx}.

The quenched case possesses only a trivial center \cite{Holland:2003jy}. This is less relevant for the dynamical case where already in QCD center symmetry is explicitly broken. The quenched case shares otherwise many similarities with quenched QCD. In particular, not withstanding the trivial center, there is an intermediate-distance linear rising Wilson potential \cite{Greensite:2006sm,Liptak:2008gx,Wellegehausen:2010ai}, like in full QCD.

The hadronic spectrum is rich in the unquenched theory \cite{Holland:2003jy,Wellegehausen:2013cya}. It is possible to form gauge-invariant states from any number and combination of quarks and antiquarks. Some of these states are necessarily hybrids \cite{Holland:2003jy}. This implies the existence of fermionic baryons \cite{Holland:2003jy,Wellegehausen:2013cya}, especially three-quark states, to be called neutrons\footnote{Actually, since there is only one flavor, it is not quite the same object as a neutron in QCD. It is called neutron to demonstrate that it is the lightest spin 1/2 three-quark state in the spectrum.}. Note that there is a fermionic baryon build up from a single quark. However, to make this state a gauge-singlet, it requires a dressing by three gluons \cite{Holland:2003jy}. As constituent gluons are usually heavier than constituent quarks, about 500 MeV, it is expected that these states become substantially heavier than the neutrons at sufficiently light quark masses. But the spectroscopy of these states is challenging \cite{Wellegehausen:2013cya}. Therefore this remains a conjecture. Note that this implies that in G$_2$-QCD also concepts like hybrid-stars, besides neutron and quark stars, could be possible.

The phase diagram of the quenched theory is similar to the one of ordinary SU(3) Yang-Mills theory. In the quenched case there is a first-order finite-temperature phase transition \cite{Greensite:2006sm,Cossu:2007dk,Bruno:2014rxa} which restores at the same temperature chiral symmetry \cite{Danzer:2008bk} and leads to a drop in topological susceptibility \cite{Ilgenfritz:2012aa,Bonati:2015uga}. In the unquenched case the full phase diagram looks quite similar to the one expected for QCD \cite{Maas:2012wr}.

In the following, we use as input the lattice data for the two different ensembles with one flavor of \cite{Wellegehausen:2013cya}, i.\ e.\ a Goldstone (diquark) mass of 247 MeV (light ensemble) and 326 MeV (heavy ensemble). At the current time these are the only ones available for which at a sufficient number of density values a reasonable statistics exist. In both cases the neutron mass is set to 938 MeV to fix the scale. This will also provide similar scales for neutron stars as for ordinary QCD. Based on the Gell-Man-Oakes-Renner relation and a constituent picture for the neutron this corresponds to about 20 and 50 MeV current quark masses, respectively. Therefore, assuming about 500 MeV for the constituent gluon mass, roughly corresponding to the QCD case, the mass of the hybrid would be estimated to be about 1.8 GeV in both cases, and thus almost twice the neutron mass. It appears therefore safe to assume that it will not play a role below a chemical potential roughly twice the neutron mass. As will be seen below, even the heaviest stable neutron star does not reach such a chemical potential at its core. This self-consistently supports us in neglecting the hybrids in the following. This exhibits a requirement in response to the first question: Full knowledge of the spectrum is necessary.

A somewhat more involved problem is the influence of the diquarks. They carry also baryon number \cite{Holland:2003jy,Wellegehausen:2013cya}, but are bosonic particles, and substantially lighter than the neutrons. However, the neutrons are stable against decay into diquarks in our case, as the only decay channel is to a diquark and a hybrid, the latter being too heavy. Since the diquarks are bosons, we assume that any amount of them will collapse, due to a lack of Fermi pressure, until a density is reached where they can be converted in a three-to-two process to neutrons. This then stabilizes against further collapse by forming a hadronic Fermi surface. Therefore, we will consider a neutron star made essentially only from neutrons in the following. Similar questions also arise with other states, including those heavier than the neutron. These are also issues concerning the first question.
 
\section{Mass-Radius relation of neutron stars}\label{s:mr}

Arguably the most characteristic feature of neutron stars is their mass-radius relation \cite{Glendenning:1997wn}. We follow here the approach of Tolman, Oppenheimer and Volkoff: We assume a static and spherically symmetric metric, and an energy-momentum tensor of an ideal fluid. This substantially simplifies the problem, but appears to be still acceptably realistic for our purpose. This yields the Tolman-Oppenheimer-Volkoff equation  \cite{Glendenning:1997wn}
\begin{equation}
\frac{dp(r)}{dr} = - \frac{[p(r)+\epsilon(r)]~[M(r)+4\pi r^3p(r)]}{r~[r-2M(r)]}\nn,
\end{equation}
\no where \(M(r)\) is the mass-energy enclosed within the radius $r$,
\begin{equation}
M(r) = 4 \pi \int_0^r \epsilon(r') ~ r'^2 ~ dr'\label{mass},
\end{equation}
\no and $p$ and $\epsilon$ are the local pressure and energy density, respectively. Thus, the equation of state is needed for the solution of this equation.

As a baseline we consider the case of non-interacting neutrons \cite{Glendenning:1997wn}. By comparison we can therefore identify the impact of the interactions. Note that by our identification of the 3-quark bound-state of G$_2$-QCD with neutrons of the same mass as the ones of QCD, the non-interacting case is identical in both theories, yielding the same results. 

For the full interacting theory, we have a problem that so far only the baryon density is available from simulations \cite{Wellegehausen:2013cya}. The energy density requires careful renormalization \cite{Kitazawa:2016dsl}, which requires to trace its development along lines-of-constant physics. Such a calculation is at the moment prohibitively expensive for G$_2$-QCD.

To make progress, we therefore need to make some ansatz for the energy density. We neglect therefore contributions from the thermal kinetic energy \cite{Kapusta:2006pm} and any other possible residual contributions \cite{Gandolfi:2013baa}. Since the rest mass is much larger than the thermal kinetic energy and the expected binding energies on the hadronic level, and the Fermi motion plays no big role inside the neutron star, we expect this to be an acceptable approximation. Of course, this will bare access to detailed quantitative statements. However, the current statistical uncertainties in the neutron mass from the lattice are at the 5\%-8\% level in our ensembles, which especially at small densities may already be larger than the neglected effects. We will return to this approximation in section \ref{s:sys}. Ultimately, this needs to be addressed by determining the energy density explicitly on the lattice.

Thus, our energy density is entirely given by the density of neutrons as a function of the baryo-chemical potential\footnote{Note that in analogy to QCD we assign the quarks a baryon number of 1/3, to create similar scales as for QCD.} $\mu$, $n(\mu)$ and their rest mass $m_n$, $\epsilon=m_n n(\mu)$. The pressure is then \cite{Kapusta:2006pm}
\begin{equation}
  p= n\mu-\epsilon=(\mu-m_n)n(\mu)\label{pressure}.
\end{equation}
\no If the baryo-chemical potential drops to the mass of the neutron, the pressure drops to zero. This signals that the surface of the neutron star is reached. Note that we assume that the neutron mass is essentially independent of the density in the relevant range. This is corroborated by the lattice results, as discussed below.

\begin{figure}
\begin{center}
\includegraphics[width=0.5\textwidth,type=pdf,ext=.pdf,read=.pdf]{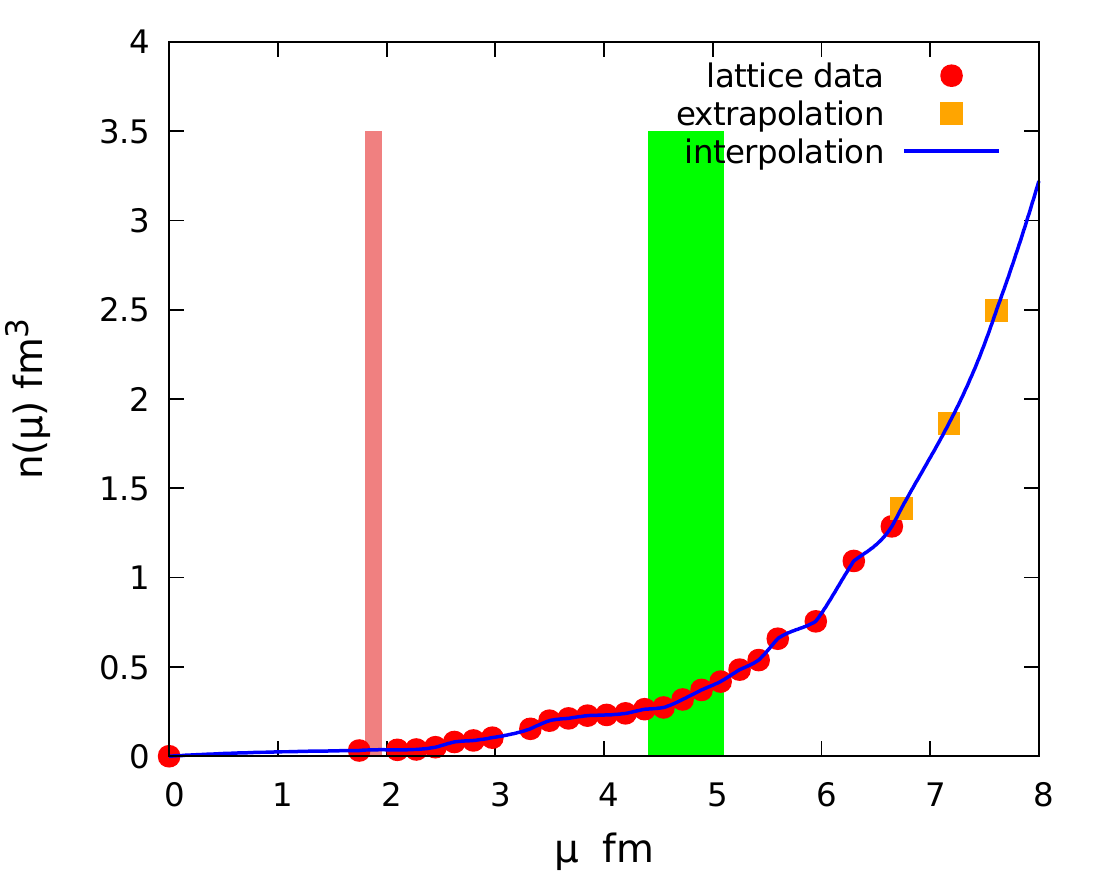}\includegraphics[width=0.5\textwidth,type=pdf,ext=.pdf,read=.pdf]{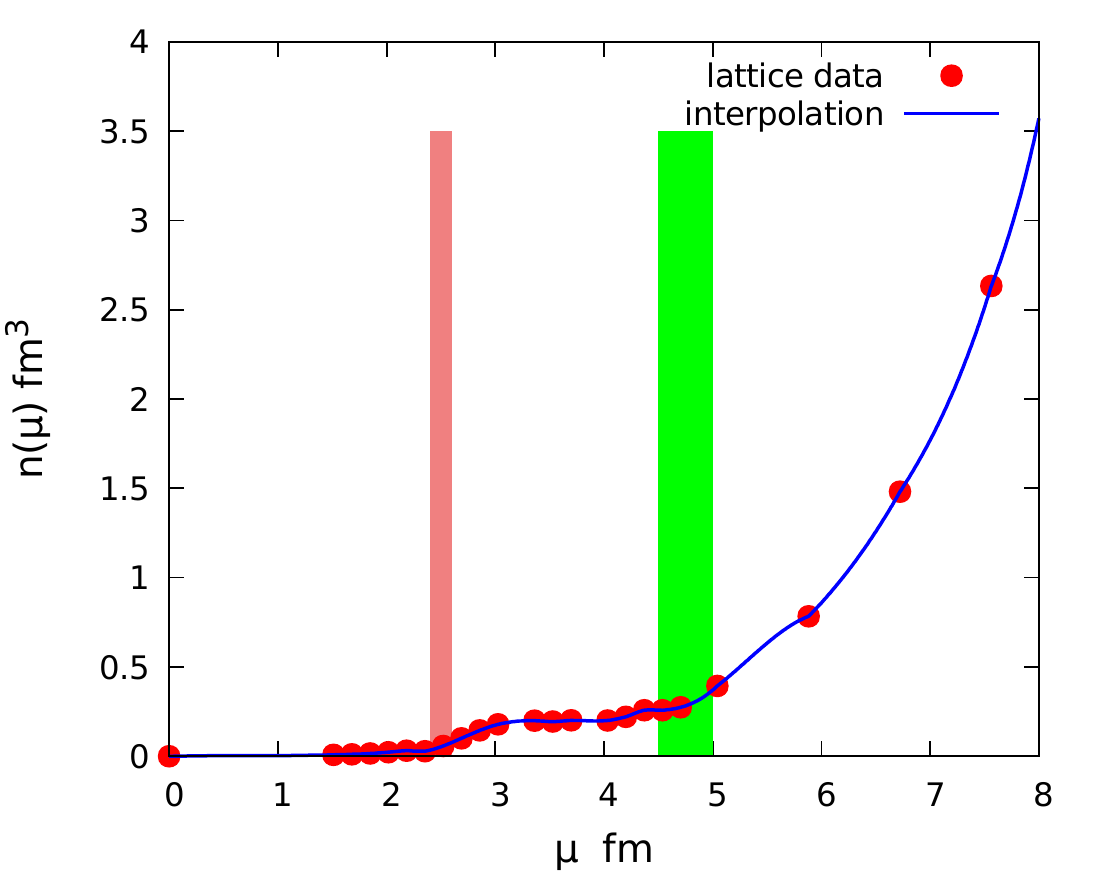} \\
\includegraphics[width=0.5\textwidth,type=pdf,ext=.pdf,read=.pdf]{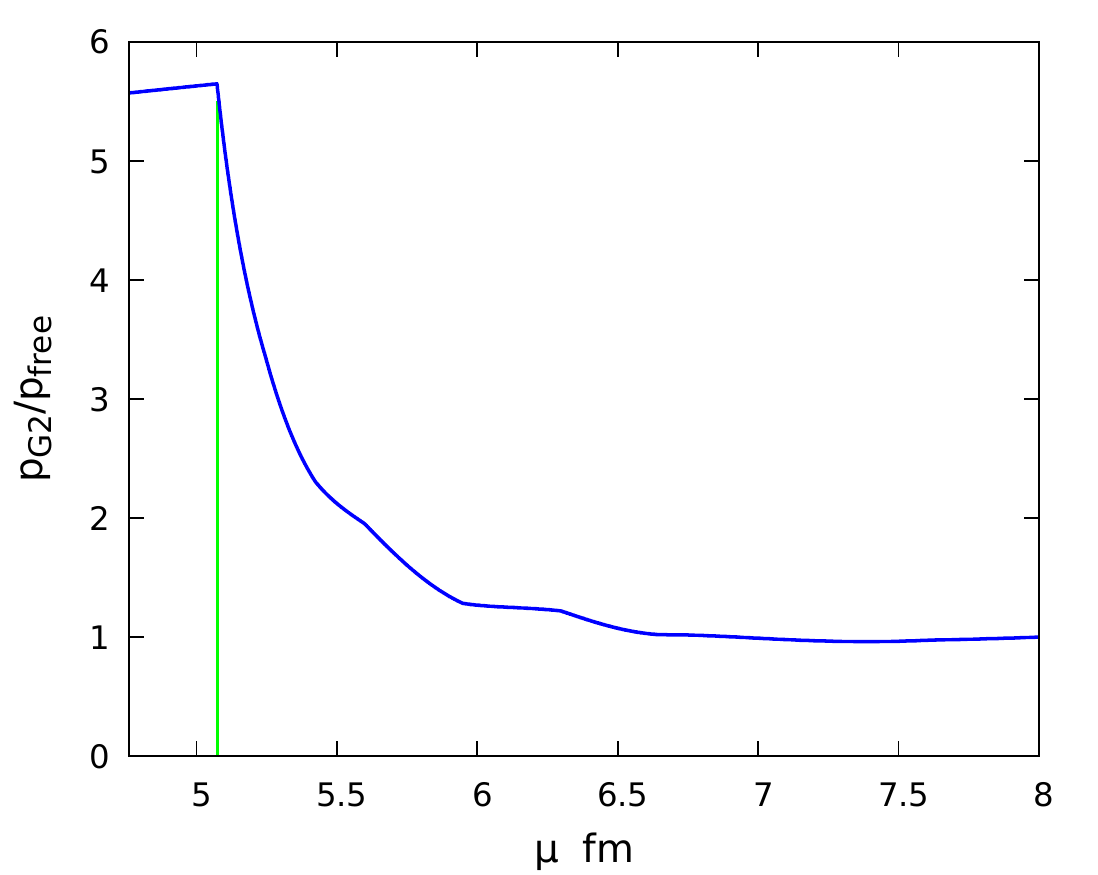}\includegraphics[width=0.5\textwidth,type=pdf,ext=.pdf,read=.pdf]{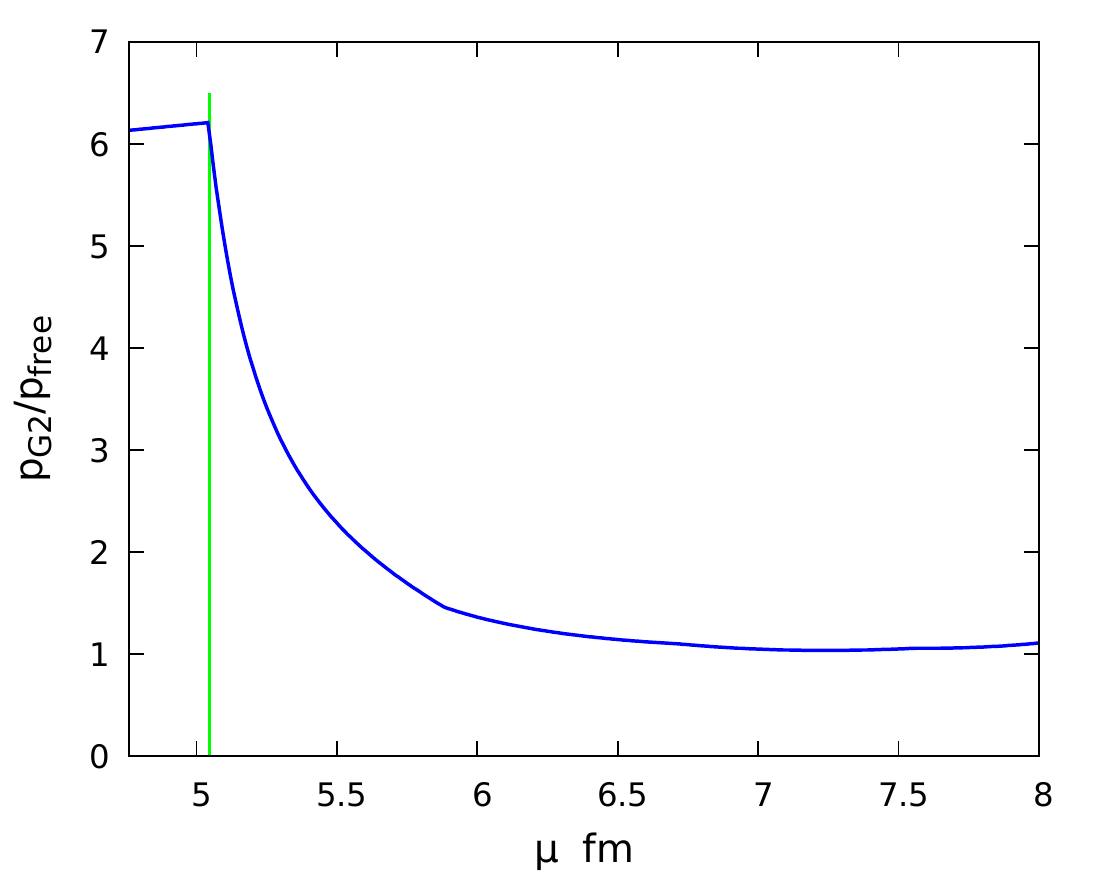}
\end{center}
\caption{\label{fig:n}The top panels show the employed baryon density as a function of baryo-chemical potential, interpolated from the lattice results, and compared to the data points, of \cite{Wellegehausen:2013cya}. The red band shows the mass of the lightest particle in the spectrum, the diquark Goldstone, while the green band is the mass of the neutron, both values also from \cite{Wellegehausen:2013cya}, both including statistical errors only. The left-hand-side shows the results for the light ensemble, the right-hand side for the heavy ensemble. The orange points in the light ensemble data are extracted from a fit, as explained in the text. Note that the statistical error bars for the lattice data are smaller than the symbol size. The actual $\mu$ occurring in the neutron stars is between $\mu_n\approx 4.8$ fm and $\mu\lesssim 8$ fm. The lower panels show the ratio between the pressure of G$_2$ QCD and the free neutron gas in the present approximation, where the green line shows the transition to the dilute neutron gas region.} 
\end{figure}

The baryon density is taken from the lattice simulations in \cite{Wellegehausen:2013cya}. Its behavior is shown in figure \ref{fig:n}. Remarkable facts are the appearance of plateaus in the density as well as that most of the transitions between plateaus happen at densities which are of the same order as the masses of the hadrons in the vacuum \cite{Wellegehausen:2013cya}. Note that this indicates that the masses of hadrons are probably not strongly affected by the density.

The two most important such transitions for the present purposes are also marked in figure \ref{fig:n}. One is the onset of a non-zero density at a scale related to the Goldstone mass, due to the silver-blaze property \cite{Cohen:2003kd}. This is the smallest chemical potential at which a non-zero density is possible at all. The second is the position where the chemical potential corresponds to the neutron mass \cite{Wellegehausen:2013cya}. This can be considered as the onset of nuclear matter. This is also the density where we assume that a hadronic Fermi surface can form, and therefore stability against gravitational collapse becomes possible. Thus, only densities above this point will be relevant. It is interesting to note that the behavior of the baryon density somewhat above this point is actually rather similar to that of a free gas of effective, heavy fermions \cite{Wellegehausen:2013cya}. This is also visible from the comparison of the pressure in our approximation to the pressure of a free neutron gas, which is also shown in figure \ref{fig:n}. The question whether such a hadronic Fermi surfaces forms cannot yet be answered using lattice simulations, and remains an open question.

Note that at very high chemical potentials lattice artifacts become dominant \cite{Maas:2012wr,Wellegehausen:2013cya}. However, it will be found that already substantially below these values the neutron stars become unstable. Therefore, this kind of lattice artifacts plays no role in the following.

The lattice only yields discrete points in the chemical potential $\mu$. We interpolated these using splines. Note that the average values of the lattice data are not necessarily strictly increasing because of statistical fluctuations. Where this was a problem for obtaining a thermodynamically consistent interpolation, we use the statistical 1$\sigma$ error range to counteract it.

For the case of the light ensemble, the lattice data at high, but still relevant, chemical potentials is too sparse for a reliable interpolation. Based on the observations in \cite{Wellegehausen:2013cya}, we use a Fermi-Dirac-type distribution $n(\mu)=n_s/(\exp(a-b\mu)+1)$, where $n_s$ is the lattice saturation density and the parameters $a$ and $b$ are fitted to the last few lattice data points, to continue the interpolation. The orange points in light ensemble data in the top-left panel in figure \ref{fig:n} are the results of the fit, while the red ones are the original lattice data.

Note that at very small chemical potentials the equation of state has to become the one of free neutrons to leading order in $\mu$ to be thermodynamically consistent \cite{Glendenning:1997wn}. Since the vacuum point is anyhow between two of the lattice data points, we enforced this by choosing as a fit-form the free-fermion behavior $c(\mu-m_n)^\frac{3}{2}$ in this range, and selected the pre-factor such that the equation of state is continuous. We will return to the systematic uncertainties induced by this in section \ref{s:sys}.

Note that the actual baryon density already starts to differ from zero at a baryo-chemical potential below the neutron mass, as the Goldstones are diquarks and also carry baryon number \cite{Wellegehausen:2013cya}. Therefore, \pref{pressure} would become negative there. We avoided this by setting the pressure to zero below $\mu=m_n$. We therefore neglect the possibility for an outer layer of pure diquark matter, for the arguments given above on the bosonic nature of the diquarks. We will reevaluate this in section \ref{s:sys}.

To check our final version of the equation of state for thermodynamic consistency, we calculated the velocity of sound. It is always below 1, consistent with thermodynamic stability.

Finally, rewriting the TOV equation using the pressure \pref{pressure} only and in terms of the baryo-chemical potential yields
\begin{equation}
\frac{d\mu(r)}{dr} = - \frac{[\mu(r)]~[M(r)+4\pi r^3p(r)]}{r~[r-2M(r)]}\label{tov}.
\end{equation}
\no To obtain the mass-radius relation we solve the coupled equations \pref{mass} and \pref{tov} for various values of the chemical potential at the core of the neutron star, starting from slightly above $m_n$. The initial conditions are obtained from the requirement of zero mass at the center of the neutron stars and the corresponding pressure value. The solution itself is possible using standard algorithms for ordinary differential equations. It is even for our interpolation numerically fully stable.

\section{G$_2$-QCD neutron star}\label{s:ns}

\begin{figure}
 \begin{center}
   \includegraphics[width=0.5\textwidth,type=pdf,ext=.pdf,read=.pdf]{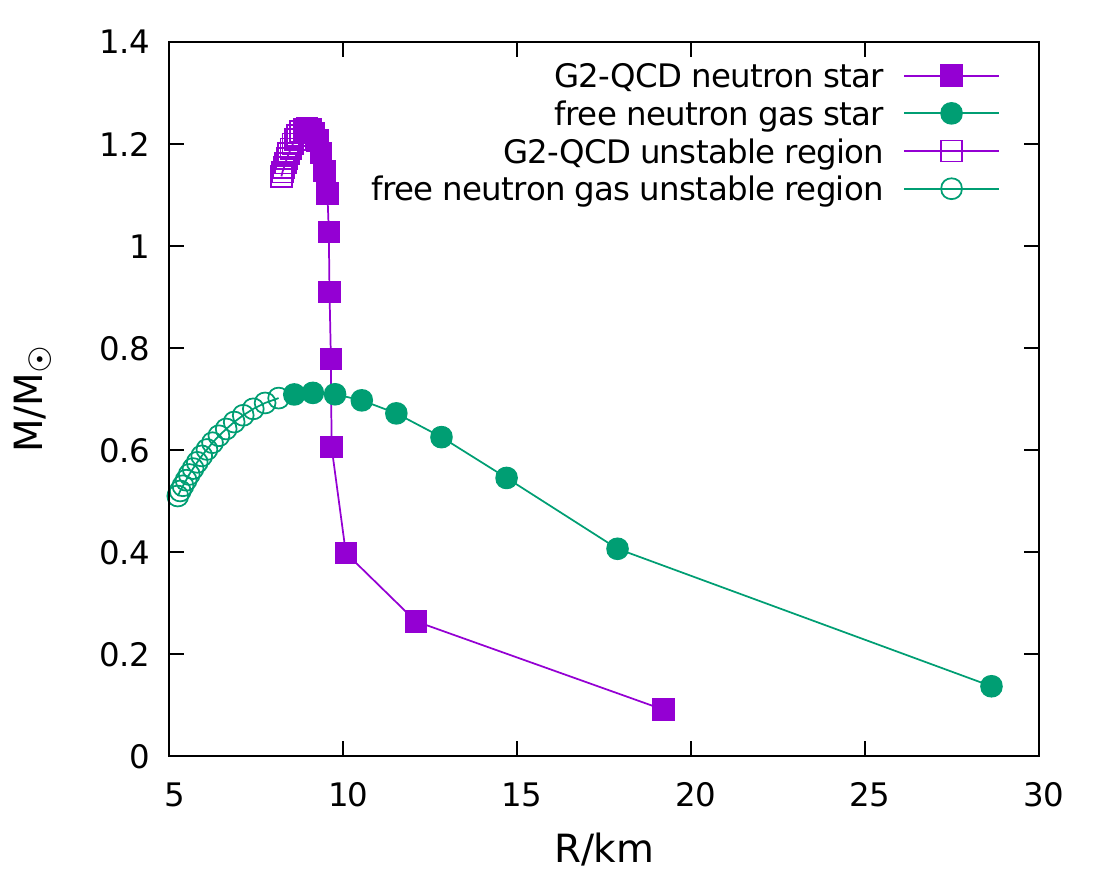}\includegraphics[width=0.5\textwidth,type=pdf,ext=.pdf,read=.pdf]{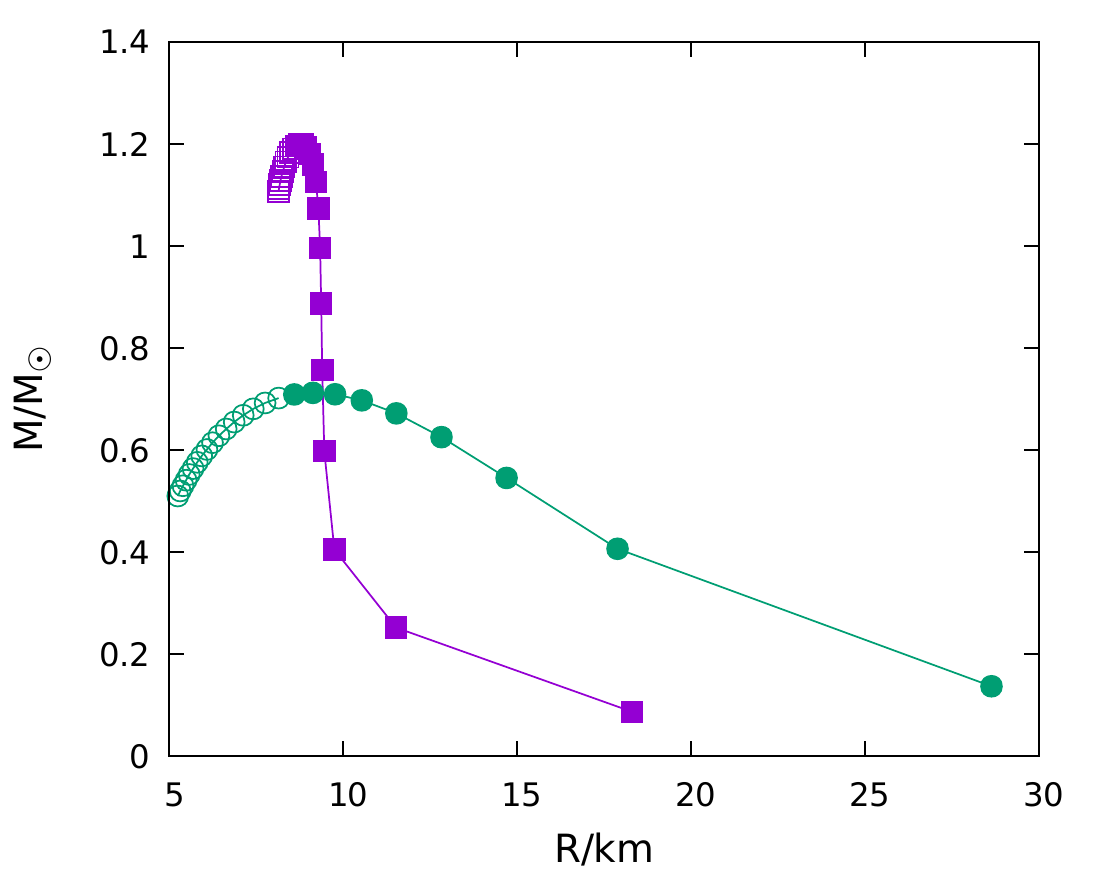}\\
   \includegraphics[width=0.5\textwidth,type=pdf,ext=.pdf,read=.pdf]{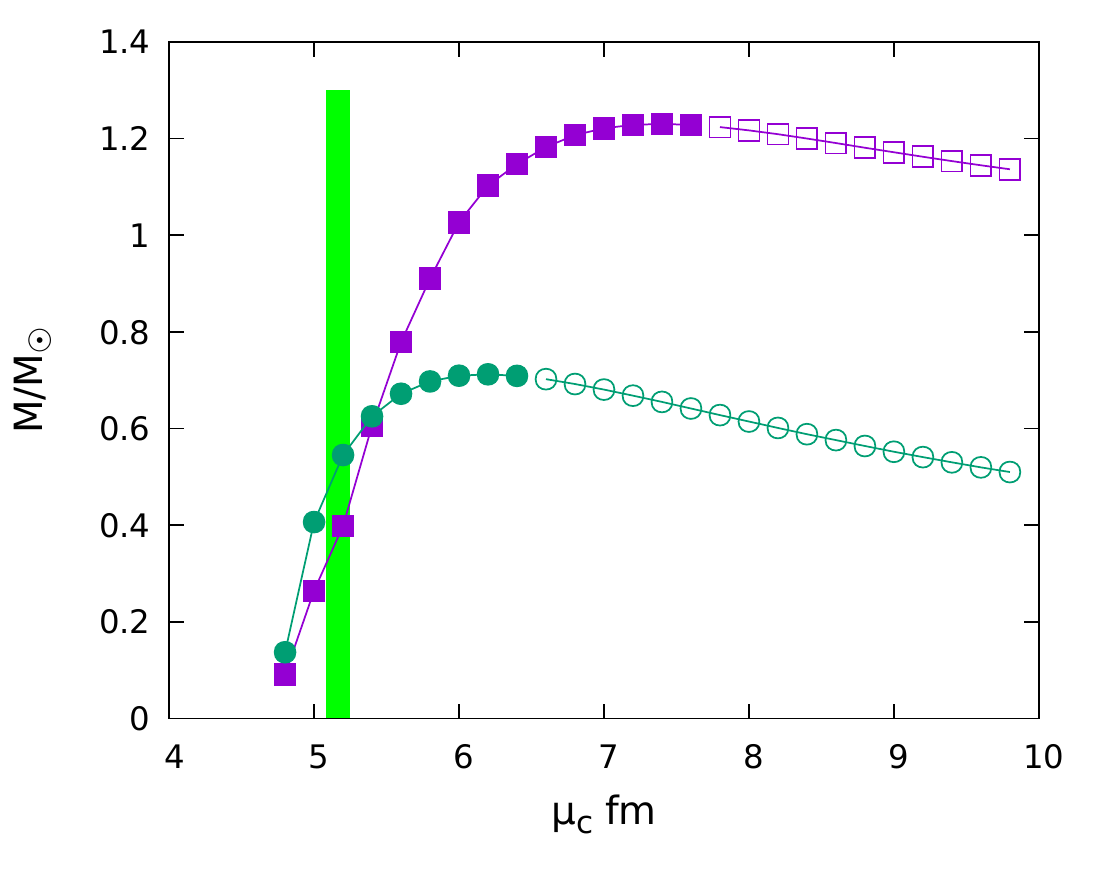}\includegraphics[width=0.5\textwidth,type=pdf,ext=.pdf,read=.pdf]{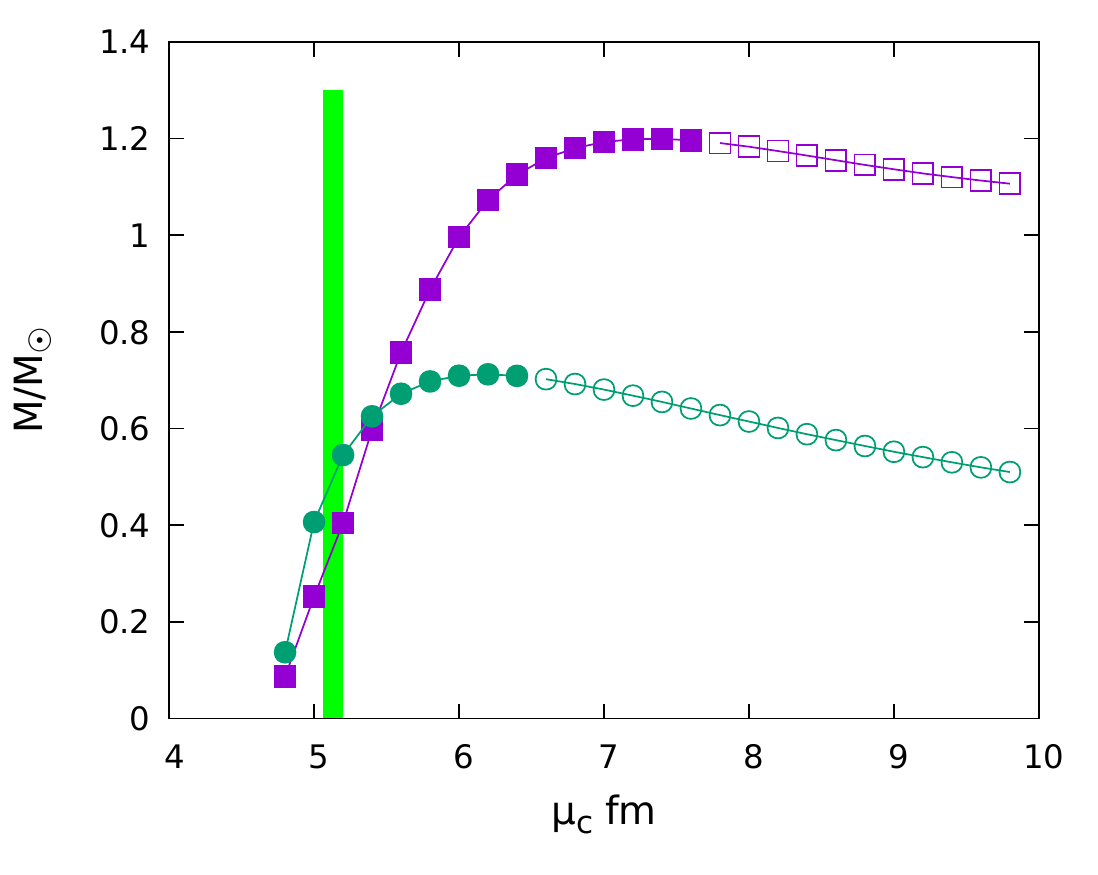}\\
   \includegraphics[width=0.5\textwidth,type=pdf,ext=.pdf,read=.pdf]{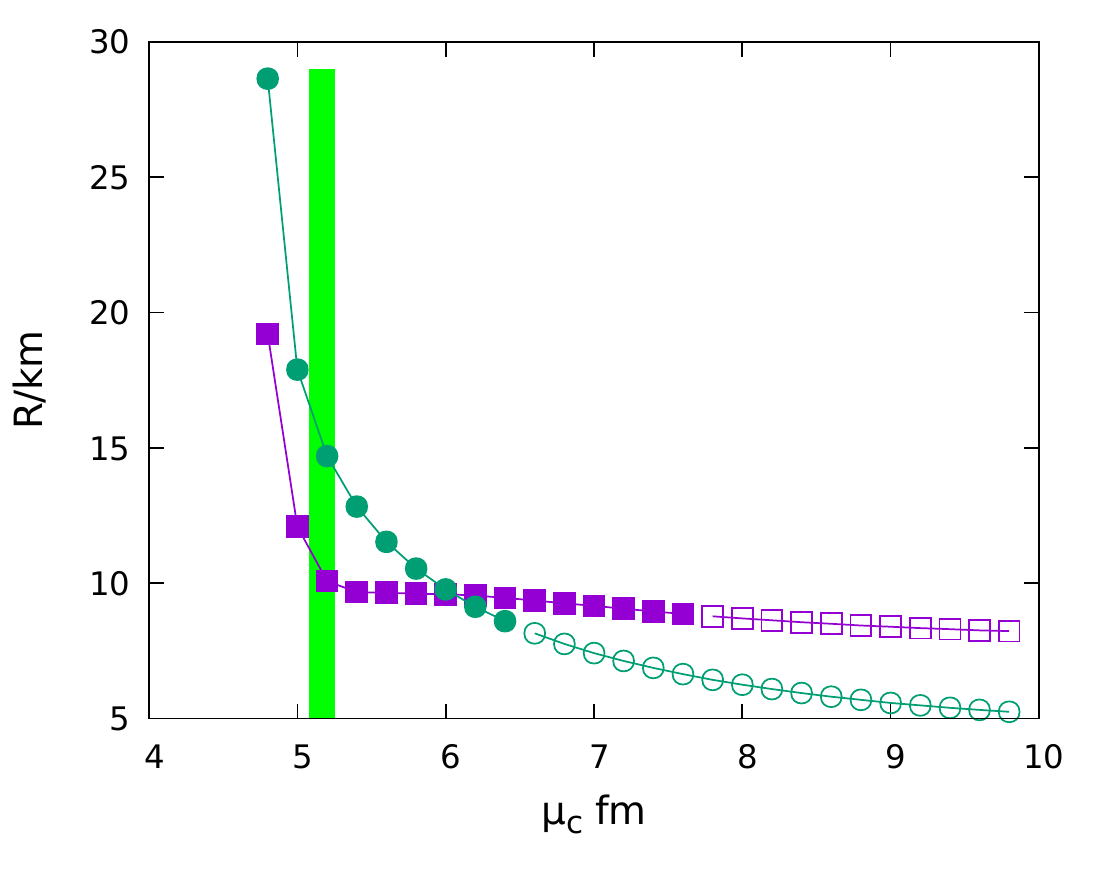}\includegraphics[width=0.5\textwidth,type=pdf,ext=.pdf,read=.pdf]{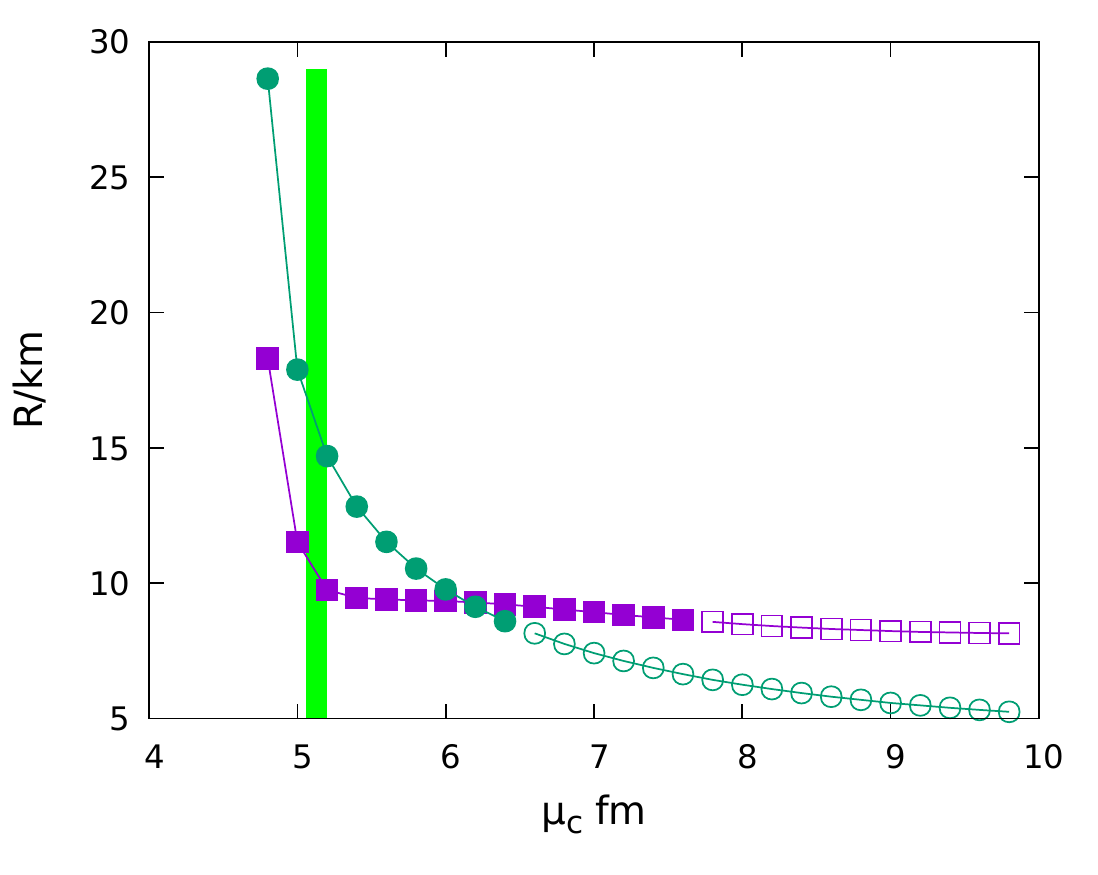}
 \end{center}
 \caption{\label{fig:mr}The top panels show the mass-radius relation for the G$_2$ neutron star (left panels: light ensemble, right panels: heavy ensemble) (purple) in comparison to the result for the free Fermi gas (green). The lower panels show the mass (middle) and radius (bottom) as functions of the central chemical potentials. The green bar in the lower panels indicate the transition point to the free gas approximation in chemical potential.}
\end{figure}

The results for the mass-radius-relation are shown in figure \ref{fig:mr}. Comparing to the free gas, the maximum mass is substantially increased. However, the maximum mass is still much lower than the maximum observed masses of neutron stars \cite {Steiner:2010fz}. Since this does not change substantially for the different Goldstone masses, and is not strongly affected by the systematic effects studied in section \ref{s:sys}, the origin of this is unclear.

Nonetheless, also for later reference, it is helpful to enumerate possible origins. One is that the situation may be quite different closer to the chiral limit. Also that only one flavor is included may play a role. And, of course, G$_2$-QCD may just be fundamentally different here than QCD. There are also possible lattice artifacts, especially from finite volume. Finally, the assumptions made to obtain the equation of state could just be incorrect. This would require to lift them. All of these issues could be checked with improved results for the equation of state from the lattice, and by moving closer in all parameters to full QCD. While too expensive at the current time, this is straightforward. Influences from $\beta$-equilibrium or other weak corrections cannot (yet) be checked in this way, and would require other approaches. All of these possible sources of influence should be kept in mind in the following. Also, the approximations made for the equation of state could be relevant for this. Some of the mentioned points will be discussed in detail in section \ref{s:sys}.

In fact, not only the maximal mass is quite similar for the heavy and light ensembles, but all quantities are very similar.

Another feature visible in figure \ref{fig:mr} is that the mass-radius relation is much narrower than in the free gas case. Thus, the neutron star is less stable against external perturbation, e.\ g.\ by accretion. Conversely, this implies that an (almost) maximum mass neutron star has essentially a unique radius.

As for conventional neutron stars, there is a mass, which cannot be exceeded. After this, the hadronic Fermi surface no longer stabilizes the star, and the star becomes unstable. The end-point of the mass-radius curve in this unstable regime is at a chemical potential where lattice artifacts slowly start to become an issue. But as the instability sets in much earlier, these lattice artifacts play no important role for stable neutron stars.

\begin{figure}
\includegraphics[width=0.5\textwidth,type=pdf,ext=.pdf,read=.pdf]{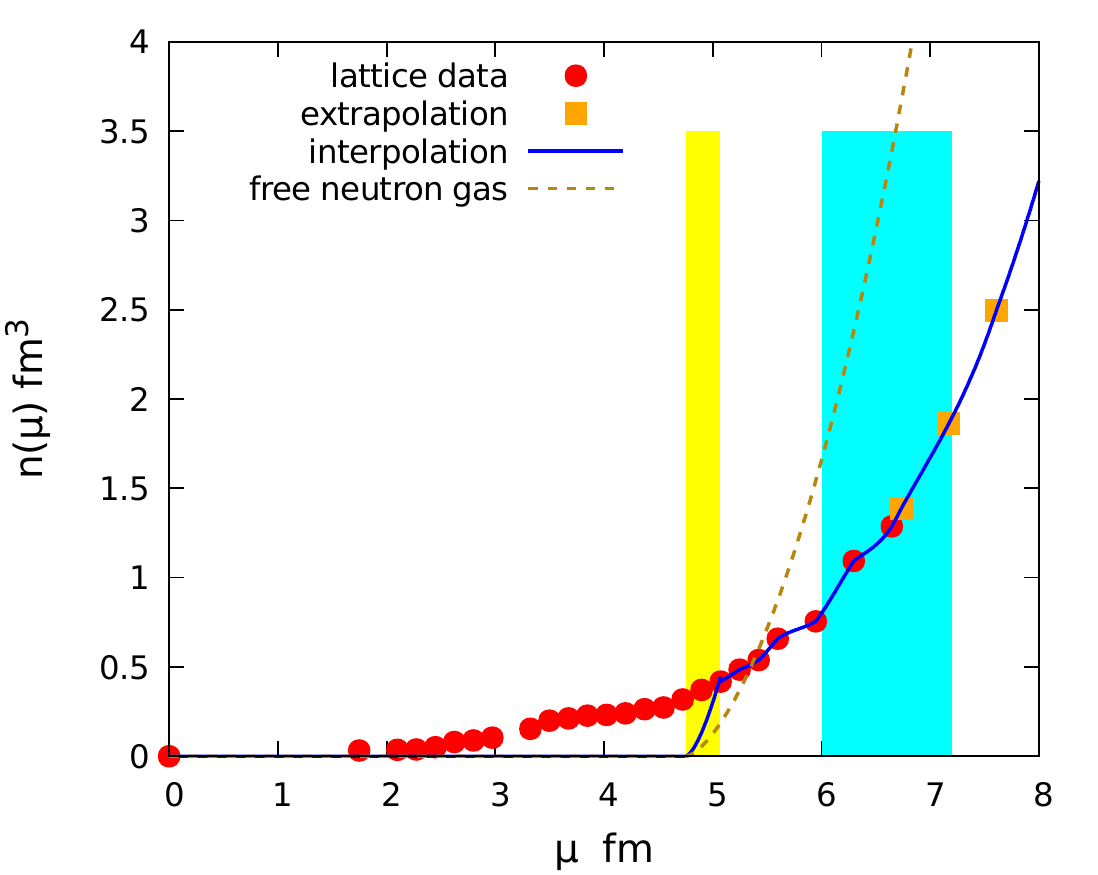}\includegraphics[width=0.5\textwidth,type=pdf,ext=.pdf,read=.pdf]{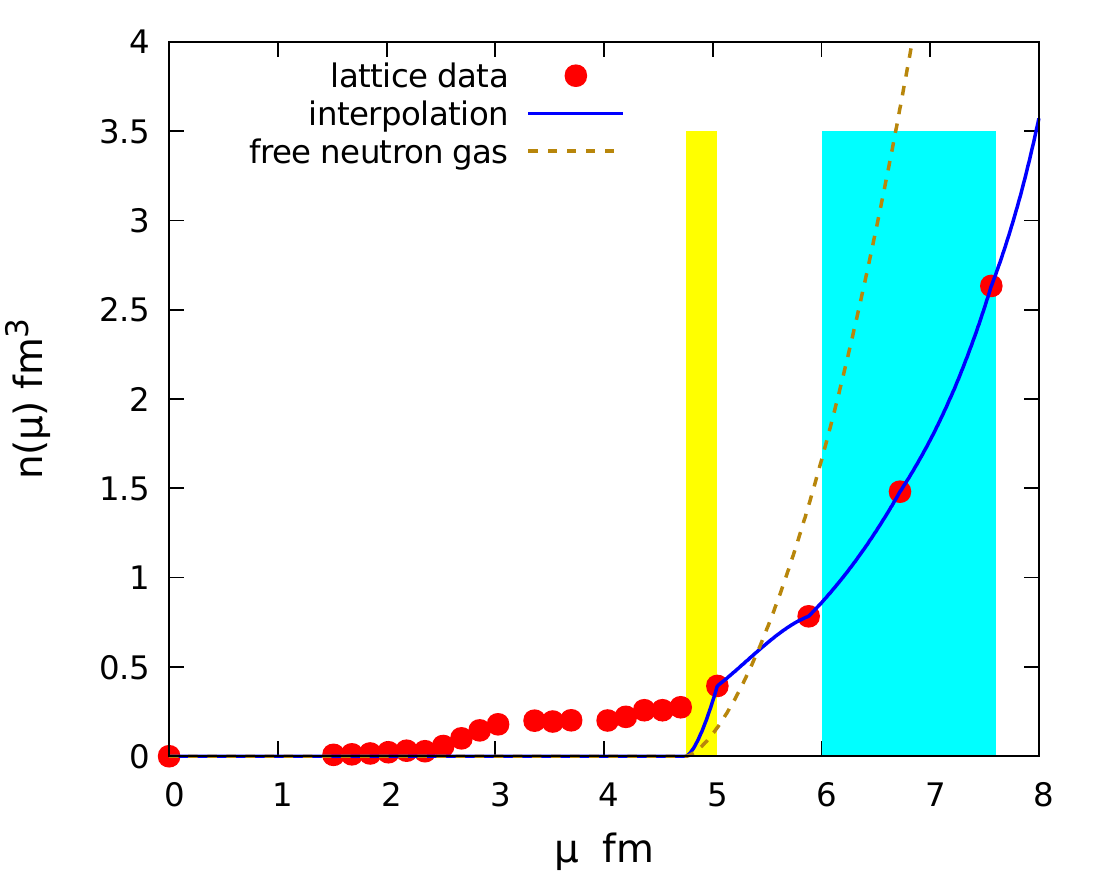}
\caption{\label{fig:redn-mu} Baryon number density vs chemical potential of nucleons. Dark golden: free Fermi gas, blue with data points: G$_2$-QCD. The yellow (left) region corresponds to the outer crust of the neutron star and the cyan (right) region the inner core of the most massive neutron stars. The left panel is for the light ensemble and the right panel for the heavy ensemble.}
\end{figure}

Finally, the chemical potentials at the center of the most massive stable neutron stars are less than twice than that of ordinary nuclear matter. That is a comparatively low value. Still, to get an idea of the scales, the actual density is roughly 2-2.5/(fm$^3$), which is about fifteen times the QCD nuclear matter density of 0.17/(fm$^3$), and almost 10 times that of G$_2$-QCD matter at the same density as nuclear matter, as can be seen from figure \ref{fig:n}.

To understand the features of the mass-radius relation, consider figure \ref{fig:redn-mu}. In this figure again the core chemical potentials for light and very massive neutron stars have been indicated. For light neutron stars, and thus large radius, the free neutron gas allows larger masses than the interacting case, opposite of what happens at larger masses. Thus, there is a crossover, where the rise for the interacting case becomes quicker. Comparing the free and the G$_2$-QCD case, it is visible that the density in the free case is smaller at smaller core chemical potentials, while this is different for heavy masses. Thus, in the interacting case the system has less mass to withstand in the core region against the collapse, and can therefore sustain more massive stars.

There is another interesting feature visible in the mass-radius relation in figure \ref{fig:mr}: At a radius of about 10 km, and a mass of 0.4 solar masses for the both ensembles, there is a strong change of slope. Comparing this to the core chemical potentials, the value is quite interesting. It is roughly the value, where the transition from the free gas to the lattice data takes place. Thus, this is where the interactions become the dominating effect, and the equation of state appears to enter a region dominated by fermionic hadrons \cite{Wellegehausen:2013cya}. Thus, this rapid change of properties impresses itself on the mass-radius relation. This could imply that changes of slope of the mass-radius relation of neutron stars could also be related to different physics. This would therefore be a highly interesting signature of observational astronomy of neutron stars. Of course, this would require rather precise radius determinations, which remain a challenge.

\begin{figure}
\includegraphics[width=0.5\textwidth,type=pdf,ext=.pdf,read=.pdf]{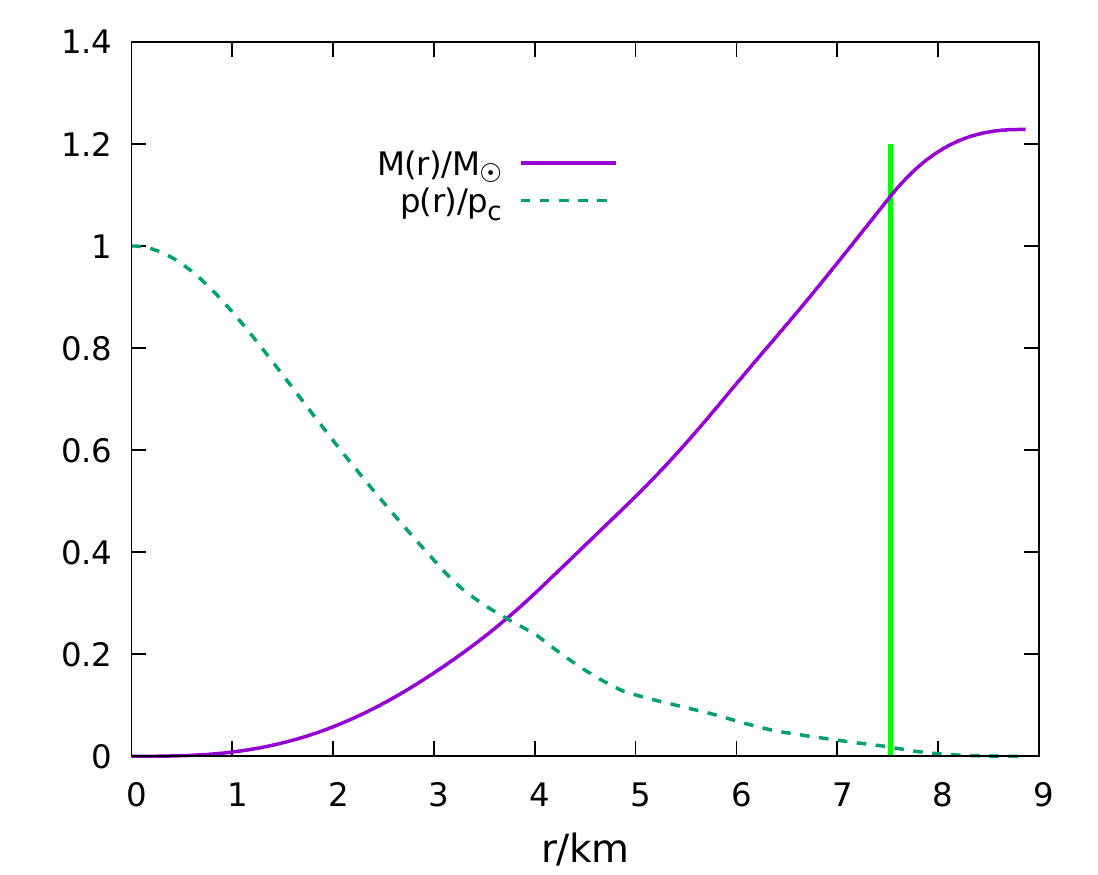}\includegraphics[width=0.5\textwidth,type=pdf,ext=.pdf,read=.pdf]{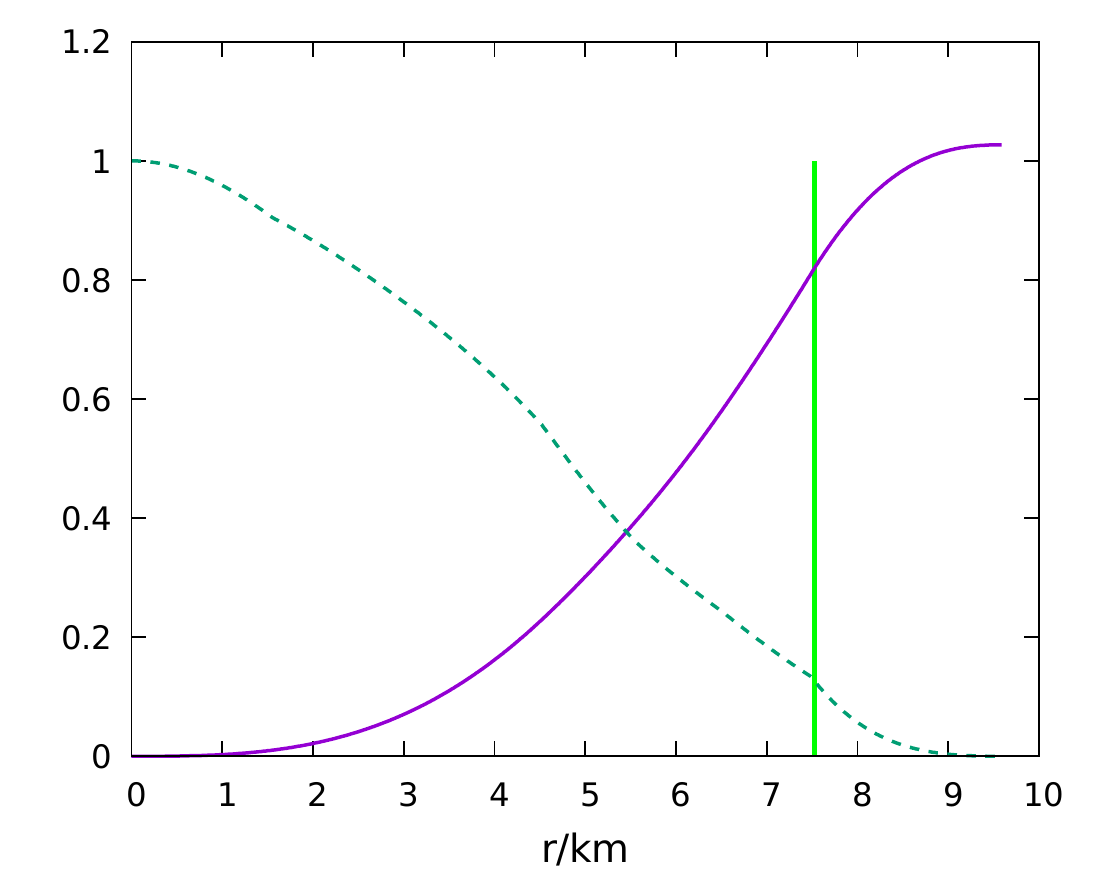}
\includegraphics[width=0.5\textwidth,type=pdf,ext=.pdf,read=.pdf]{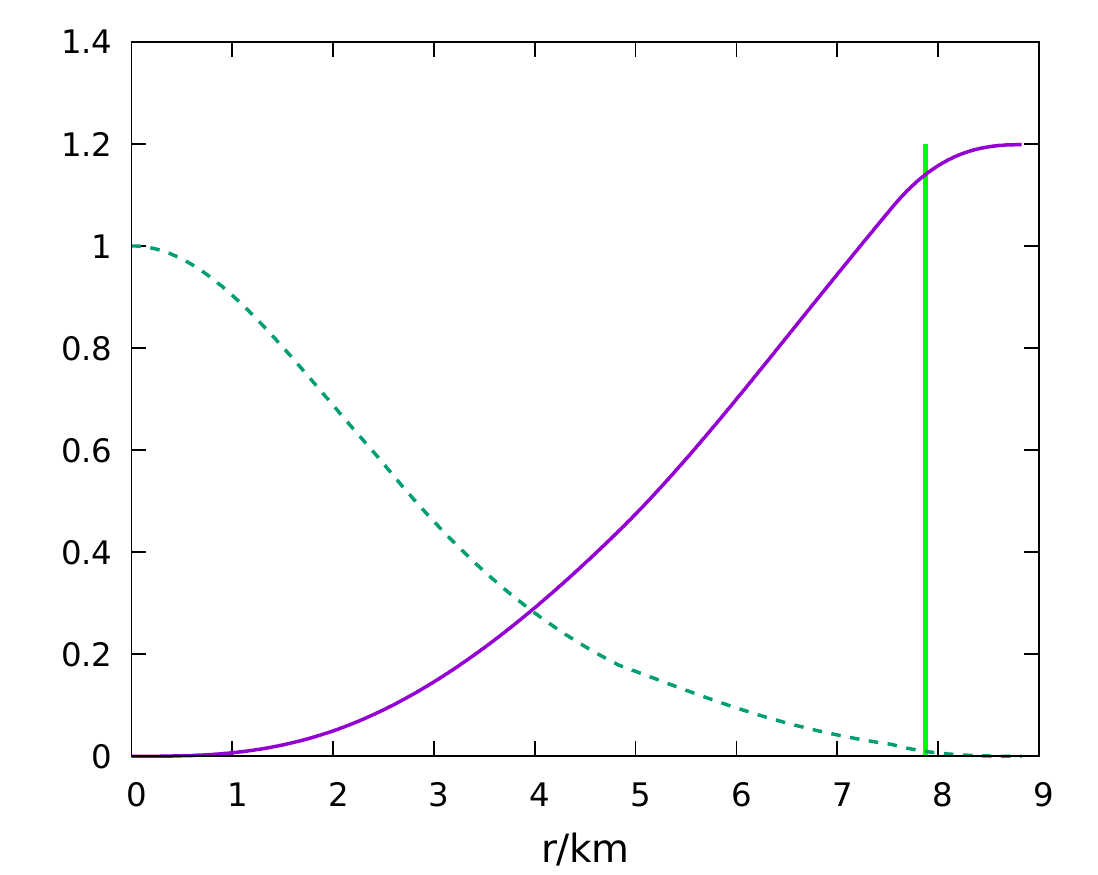}\includegraphics[width=0.5\textwidth,type=pdf,ext=.pdf,read=.pdf]{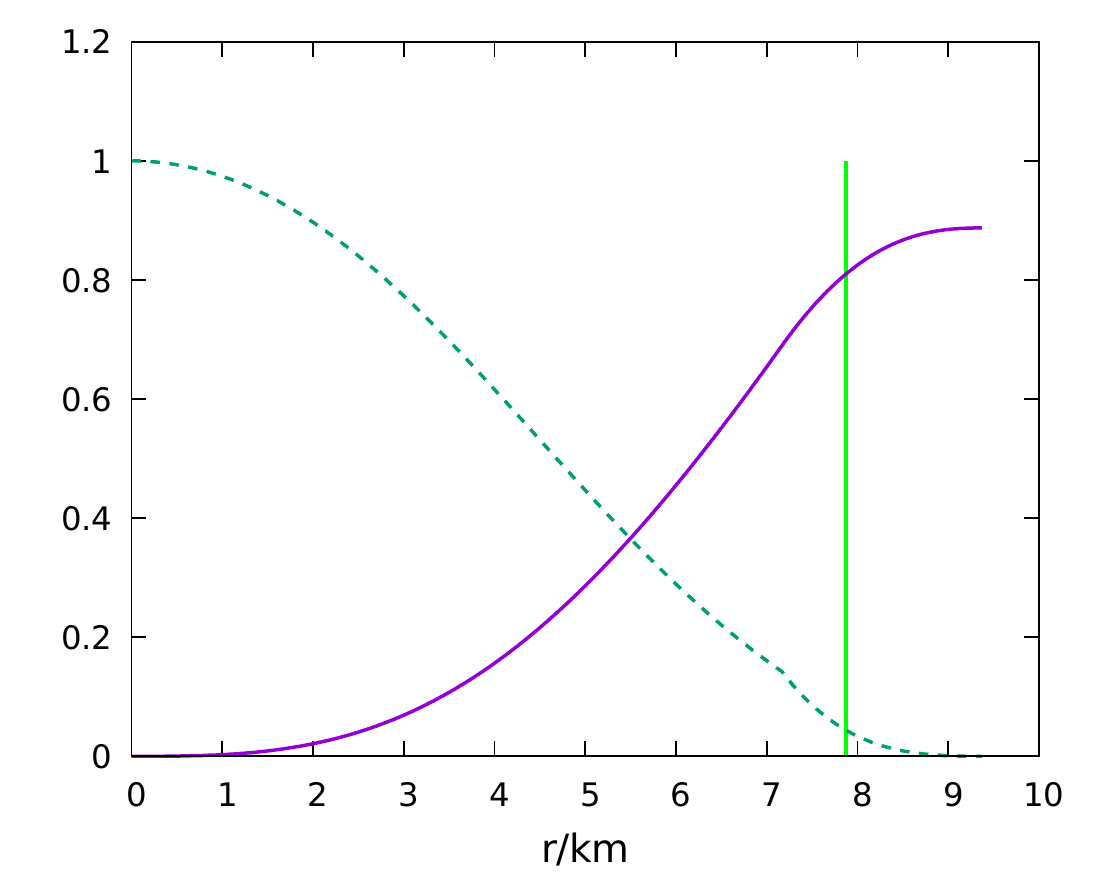}
\caption{\label{fig:profile}Accumulated mass in solar mass unit (purple) and ratio of the pressure to the central pressure (green) as a function of distance from the center of the neutron star for two different central chemical potentials. The top panels are for the light ensemble at $\mu_c=7.6$ fm$^{-1}$ (left) and $\mu_c=6$  fm$^{-1}$ (right), and the bottom panel for the heavy ensemble at $\mu_c=7.2$ fm$^{-1}$ (left) and $\mu_c=5.8$ fm$^{-1}$ (right), respectively. The green line indicates the transition to the free gas in the chemical potential.}
\end{figure}

\begin{figure}
\begin{center}
\includegraphics[width=0.5\textwidth,type=pdf,ext=.pdf,read=.pdf]{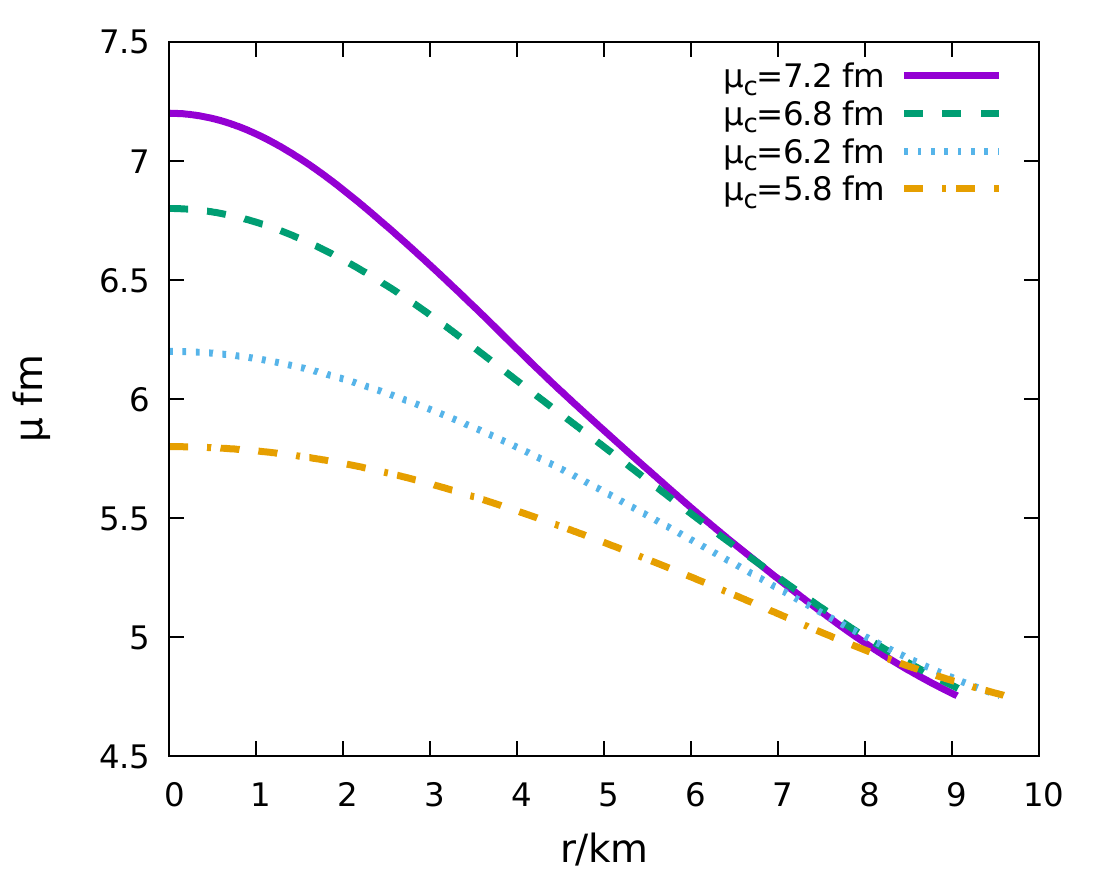}\includegraphics[width=0.5\textwidth,type=pdf,ext=.pdf,read=.pdf]{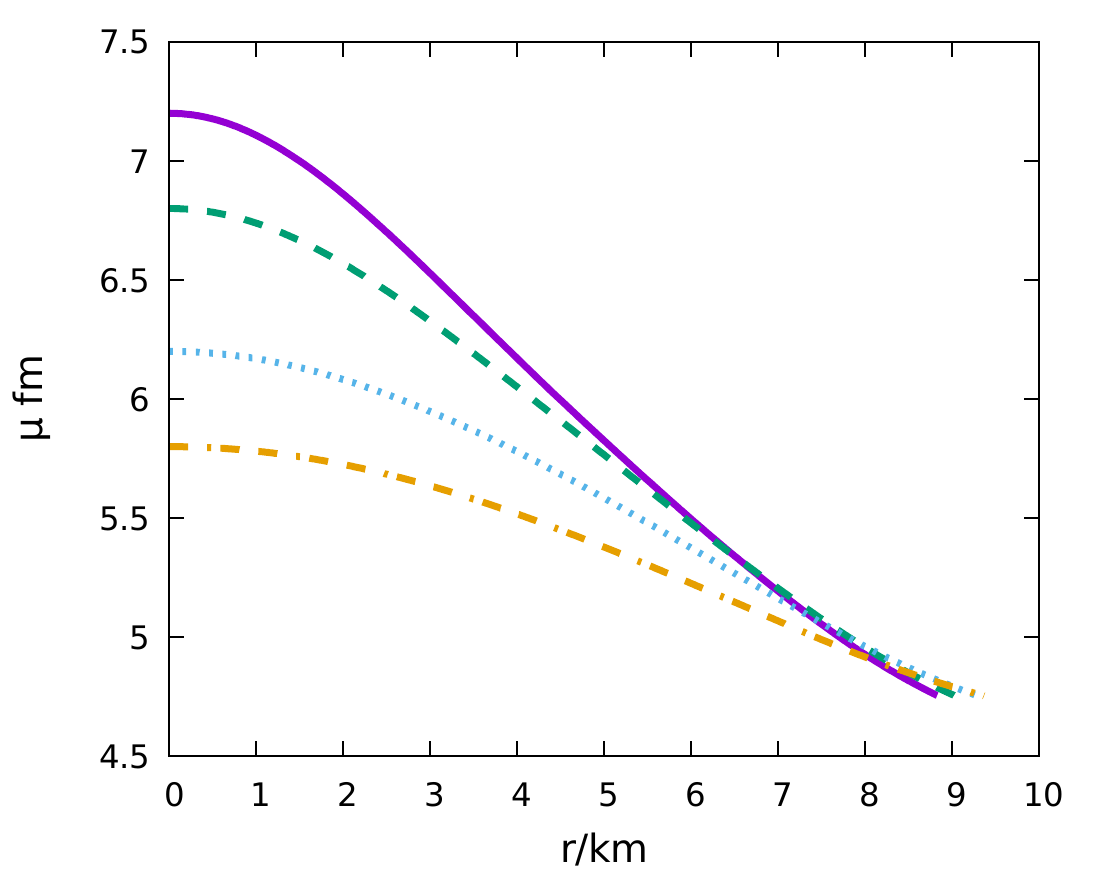}
\includegraphics[width=0.5\textwidth,type=pdf,ext=.pdf,read=.pdf]{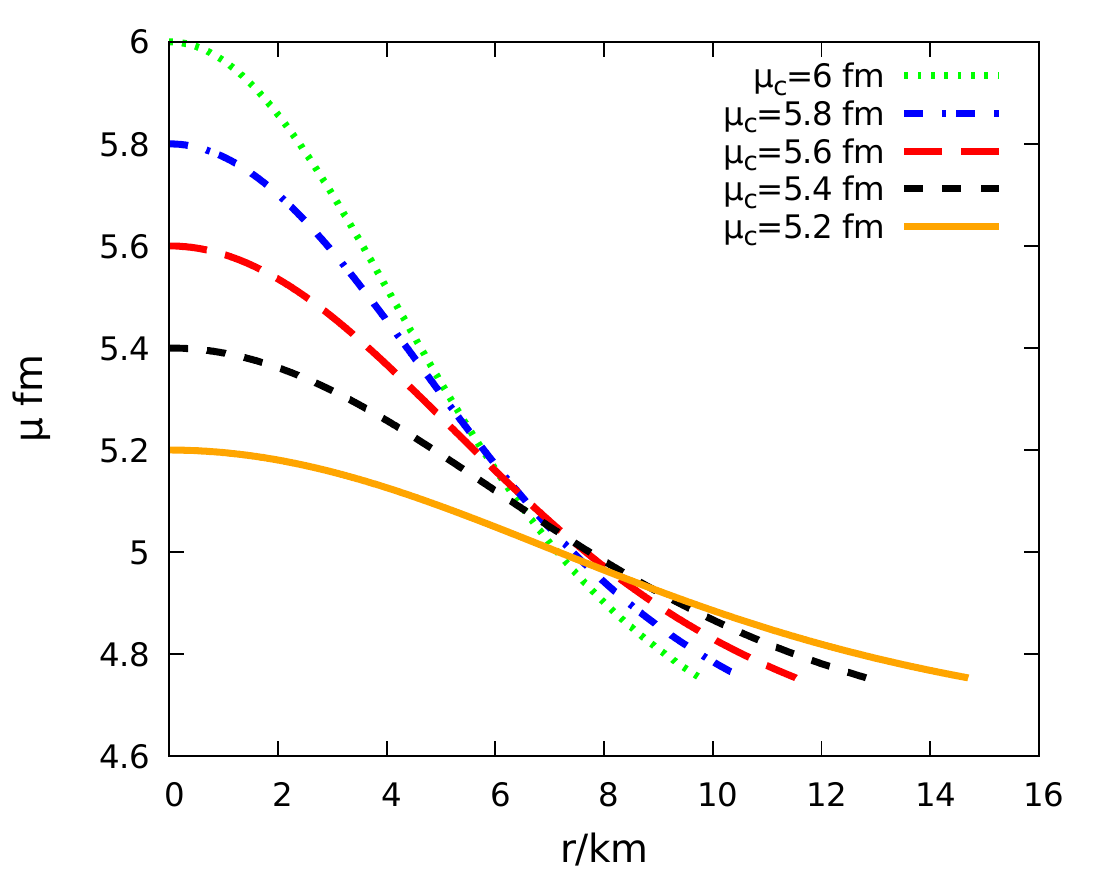}
\end{center}
\caption{\label{fig:mu}The chemical potential as a function of distance from the center for different values of the central chemical potential for G$_2$-QCD (top left: light ensemble, top right: heavy ensemble) and for the free neutron gas (bottom). Note that the slope is negative throughout, as is required for a star in stable equilibrium.}
\end{figure}

It is therefore also interesting to look inside the G$_2$-QCD neutron star to see, whether in their profile the different regions are distinguished. This is studied in figure \ref{fig:profile} for the mass and pressure and in figure \ref{fig:mu} for the chemical potential. In fact, the change is seen also in the mass profile. The point of the last step, and thus the change in slope, is roughly at the same place where the mass profile has an inflection point. The pressure or chemical potential profile does not show a similar behavior.

By comparing the plots of figure \ref{fig:profile} we observe also another interesting feature. The vertical green bar in figure \ref{fig:profile} indicates the point where we reach the dilute gas domain of our equation of state. At lower central chemical potential the amount of mass and the relative size of the outer shell, which is essentially a non-interacting neutron gas, increases. This is a consequence of the TOV equation, as here less central gravitational pull is present, and therefore the outer crust gets less compressed.

In other words, the strongly-interacting region becomes smaller when lowering $\mu_c$. This feature starts to become stronger after the change of slope, i.\ e.\ for neutron stars with a radius above roughly $8$ km. 

This is also visible when studying the chemical potential profiles in figure \ref{fig:mu}. The interacting case supports much larger central chemical potentials than the non-interacting case. Also the rate of change is quicker for the interacting case than for the free case.

We note, however, that this implies that the region we approximate as a free gas is relevant. Fortunately, this has over a reasonable range little qualitative influence on the results, but some quantitative one. Therefore the discussion above remains true within the systematic uncertainties of the present investigation. These uncertainties will now be discussed in detail in the following section.

\section{Estimates of systematic error sources}\label{s:sys}

In the following various systematic effects from the necessary approximations made in the main text to use the lattice data as the only input are investigated. Besides these effects also other systematic effects can influence our results, especially due to the finite volume and discretization errors in the lattice calculations. However, as long as no better lattice data are available, there is no possibility to estimate these effects.

\subsection{Effects of the interface point to the free gas}

As we noted, for the chemical potentials relevant close to the surface no lattice data point is available. We therefore extrapolated the lattice data starting somewhat below the last data point included in the interpolation towards smaller chemical potentials, and then fitted a free-gas ansatz for the remainder region of the chemical potential until $\mu=m_n$. The transition point in chemical potential from extrapolation to ansatz is called $\mu_t$ in the following. We note that there are lattice points available below $\mu=m_n$ with non-zero baryon density. However, in this region the baryon density is generated essentially only by the bosonic Goldstones, which cannot form a hadronic Fermi surface. Using therefore an interpolation including also the data points at even lower chemical potentials, we expect that more and more of the baryon density would be created from the 'wrong' degrees of freedom, obscuring the physical behavior. We will come back to this point.

\begin{figure}
\center
\includegraphics[width=0.5\textwidth,type=pdf,ext=.pdf,read=.pdf]{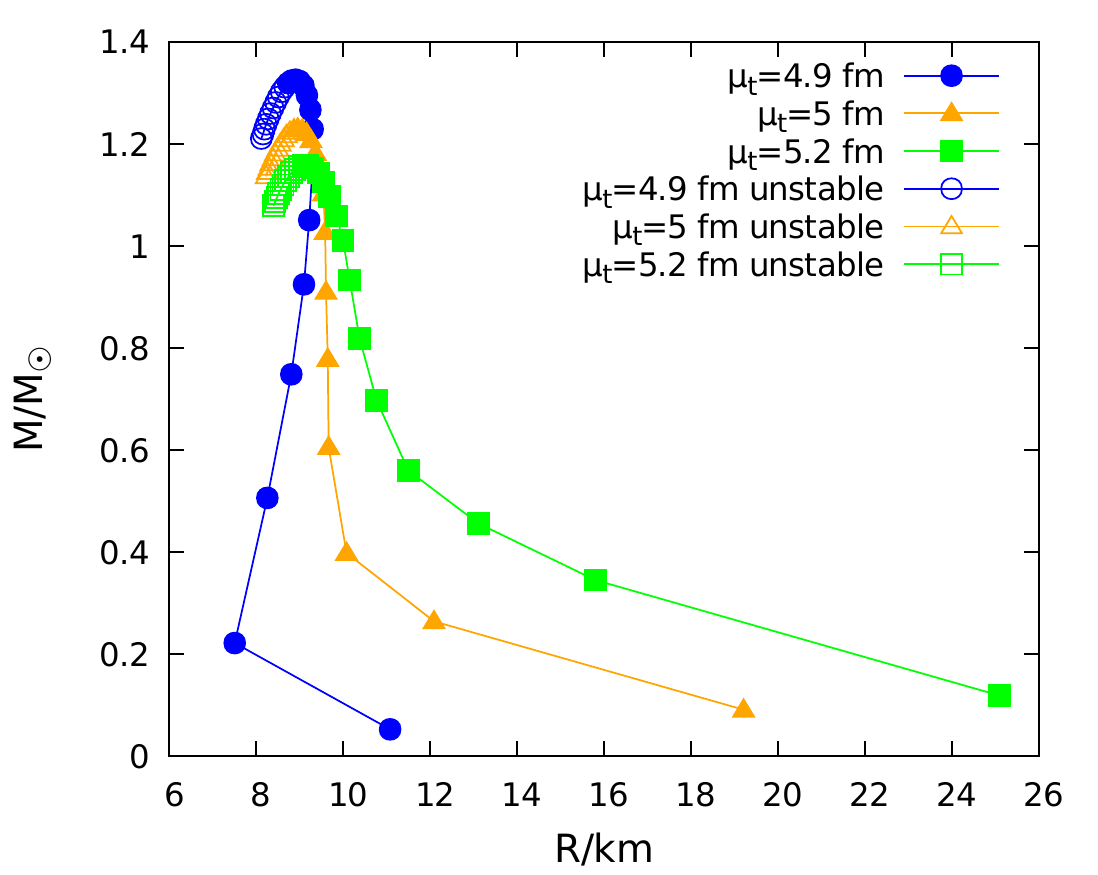}\includegraphics[width=0.5\textwidth,type=pdf,ext=.pdf,read=.pdf]{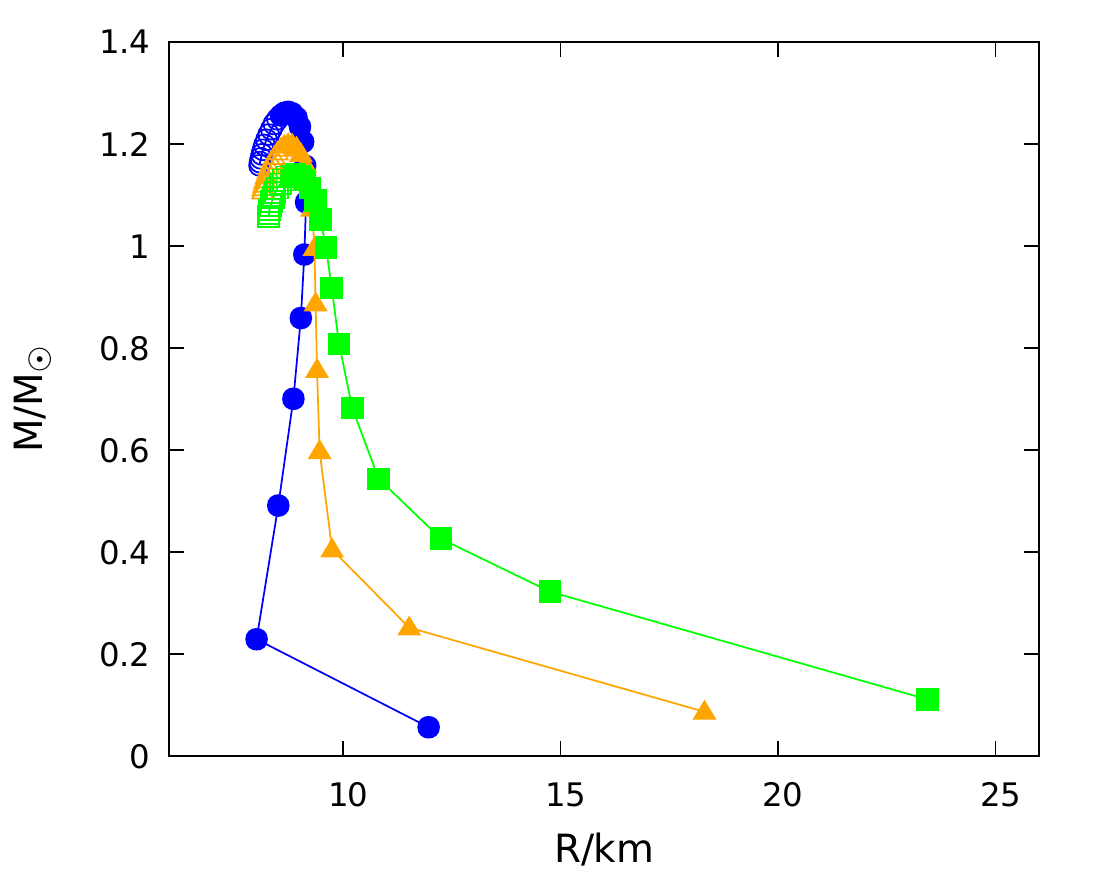}
\caption{\label{fig:M-Rt}Mass-radius curves for three different values of the chemical potential of transition to the dilute Fermi gas region, $ \mu_t=4.9$ fm$^{-1}$ (blue curve) $\mu_t= 5$ fm$^{-1}$ (orange curve), and $\mu_t= 5.2$ fm$^{-1}$ (green curve) for the light ensemble (left panel) and the heavy ensemble (right panel)(lines connecting numerical points are to guide the eye).}
\end{figure}

The result of varying the transition point $\mu_t$ is shown in figure \ref{fig:M-Rt}. The maximum mass and the corresponding radius are only weakly affected, as are other features characterizing the most massive stars. What is more strongly affected are the low-mass stars. Especially, somewhere below $\mu_t=5$ fm$^{-1}$, but still above $\mu_t=m_n\approx 4.8$ fm$^{-1}$ the bending at low masses changes, and light neutron stars become unstable\footnote{We assume here that matter at these chemical potentials behaves like ordinary matter, as the so-called self-bound matter could still stabilize such a star.}. This is a direct consequence of the violation of thermodynamic consistency \cite{Glendenning:1997wn}. On the other hand, going close to the lowest lattice point in $\mu_t$ makes heavier stars larger, and low-mass stars closer in behavior to the free gas. This also reduces how strong the impact of the change between the free case and the interacting case on the slope is, washing out, but not eliminating, the effect.

\begin{figure}
\includegraphics[width=0.50\textwidth,type=pdf,ext=.pdf,read=.pdf]{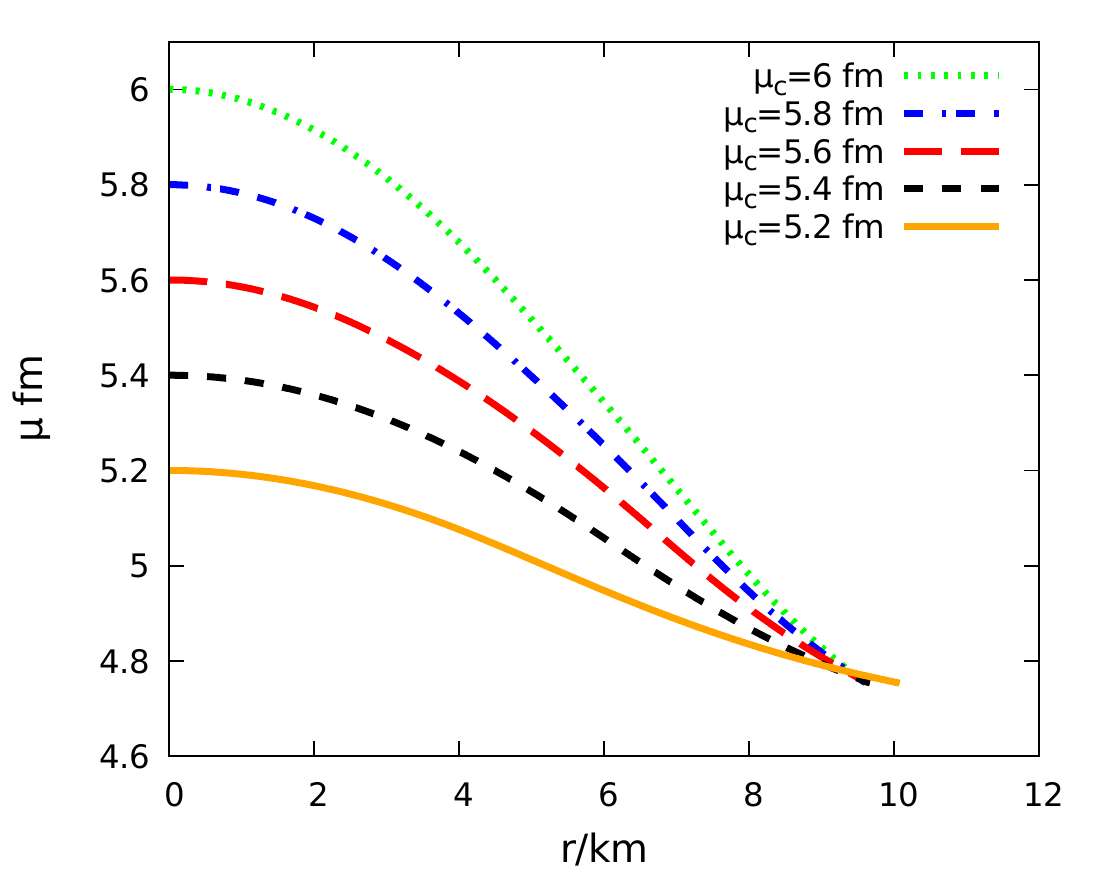}\includegraphics[width=0.50\textwidth,type=pdf,ext=.pdf,read=.pdf]{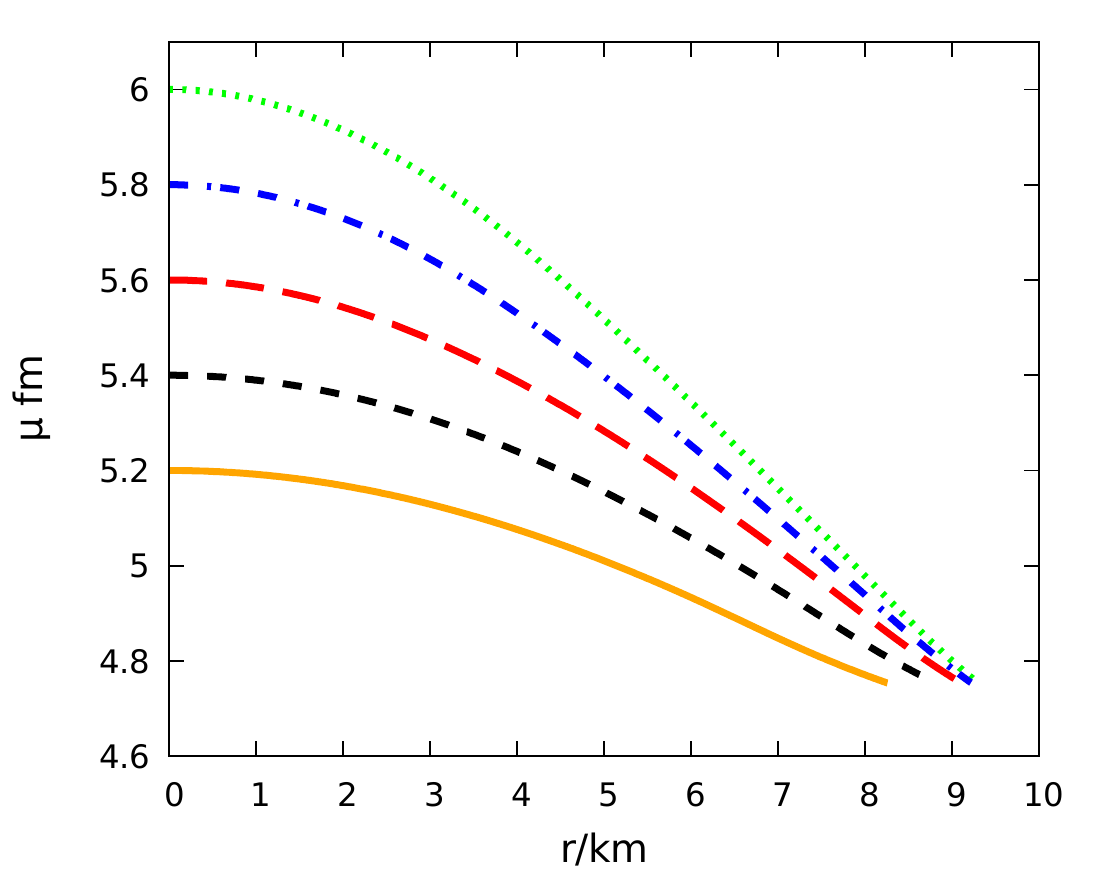}\\
\includegraphics[width=0.50\textwidth,type=pdf,ext=.pdf,read=.pdf]{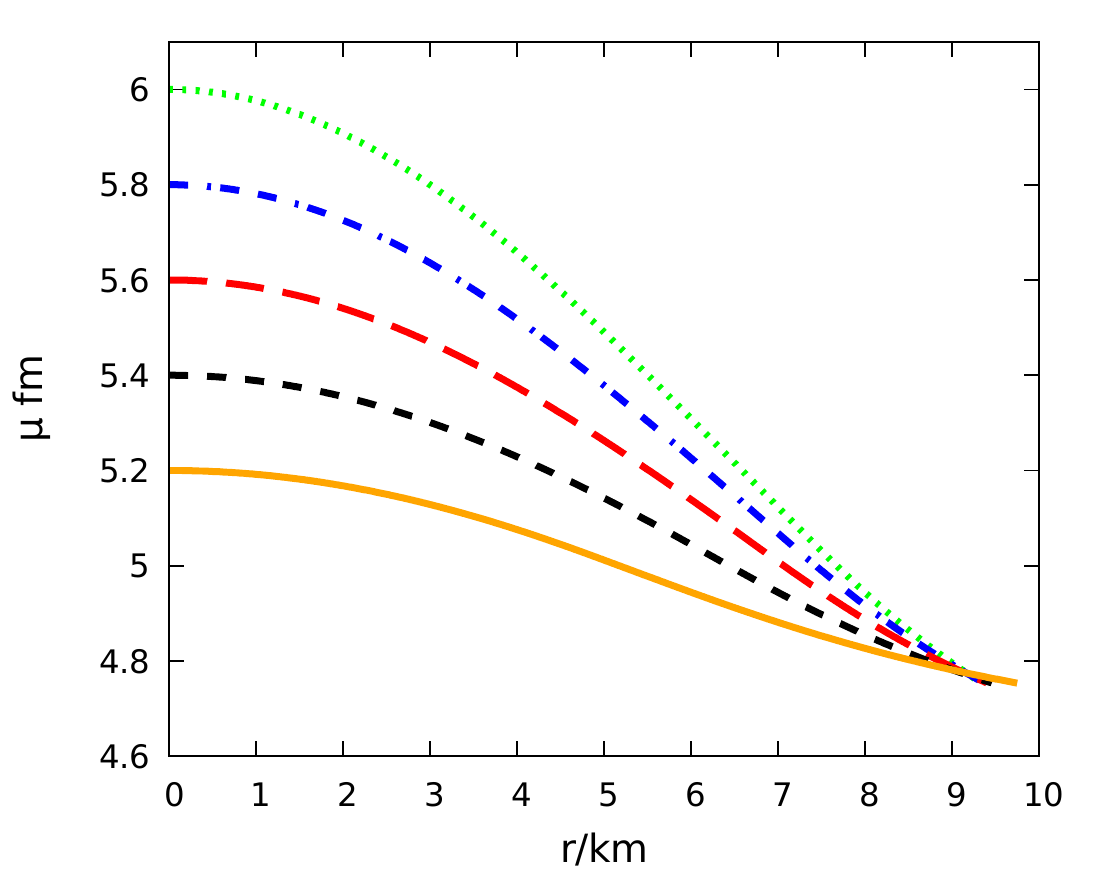}\includegraphics[width=0.50\textwidth,type=pdf,ext=.pdf,read=.pdf]{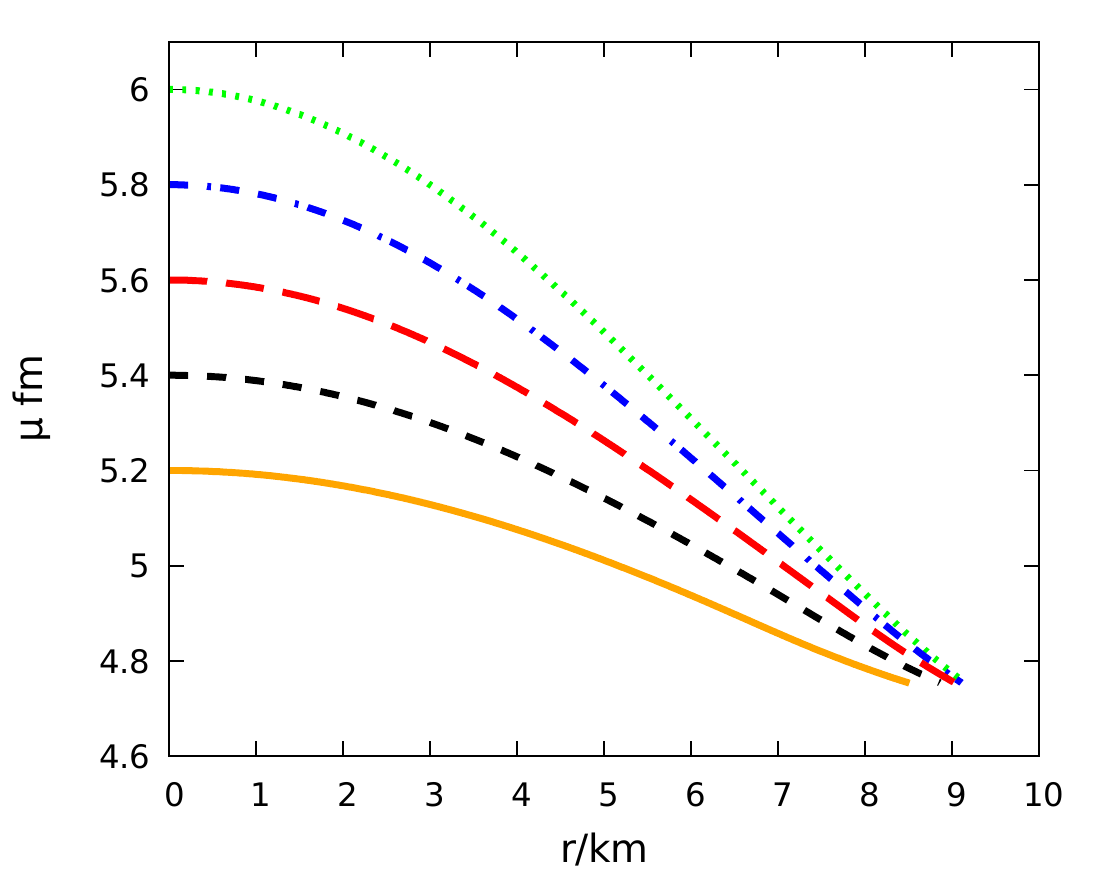}
\caption{\label{fig:mu-rt} $\mu(r)$ inside the star for $\mu_t=5$ fm$^{-1}$ (left panel) and $\mu_t=4.9$ fm$^{-1}$ (right panel). The top panels are for the light ensemble and the bottom panels for the heavy ensemble.}
\end{figure}

As can be seen from figure \ref{fig:mu-rt}, the appearance of the instability is also signaled by the fact that the density profiles no longer cross for different central chemical potentials. In this case the star therefore becomes unstable as a reduction in chemical potential is possible by shrinking.

As noted above, the baryon density is stronger influenced by the Goldstones when decreasing the chemical potential, of which we do not expect a stable star. This is in agreement with the observation. Thus, we need to exclude this effect, giving us a lowest $\mu_t$ for which still all stars are stable. On the other hand, at larger $\mu_t$, we discard substantial amount of interaction effects, as in this region the density almost doubles, see figure \ref{fig:n}.

We therefore require $\mu_t$ to be such that all neutron stars lighter than the most massive ones are stable. At the same time, we move close to the critical $\mu_t$ in the main text, to include as much as possible from the interaction effects. Thus we choose $\mu_t=5.07$ fm$^{-1}$ for the light ensemble and $\mu_t=5.04$ fm$^{-1}$ for the heavy ensemble in the main text.

\subsection{Effects of the interpolation}

\begin{figure}
\center
\includegraphics[width=0.5\textwidth,type=pdf,ext=.pdf,read=.pdf]{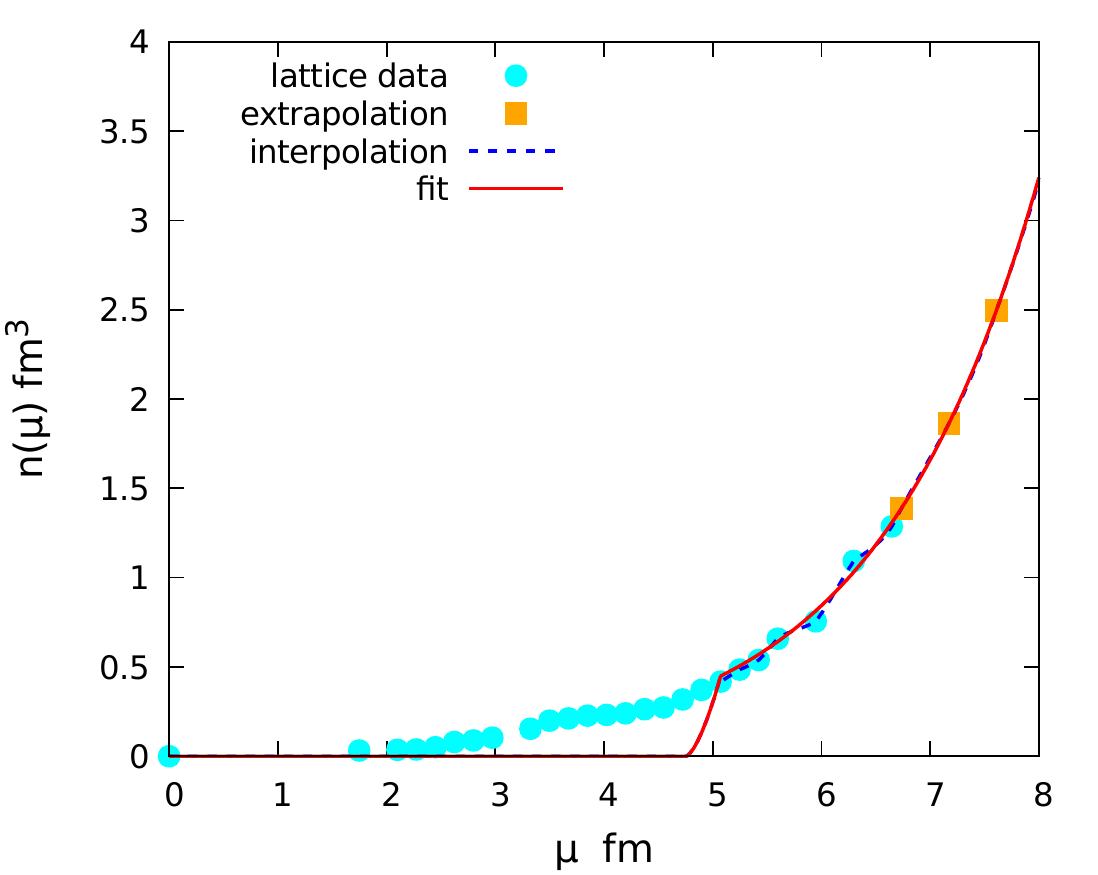}\includegraphics[width=0.5\textwidth,type=pdf,ext=.pdf,read=.pdf]{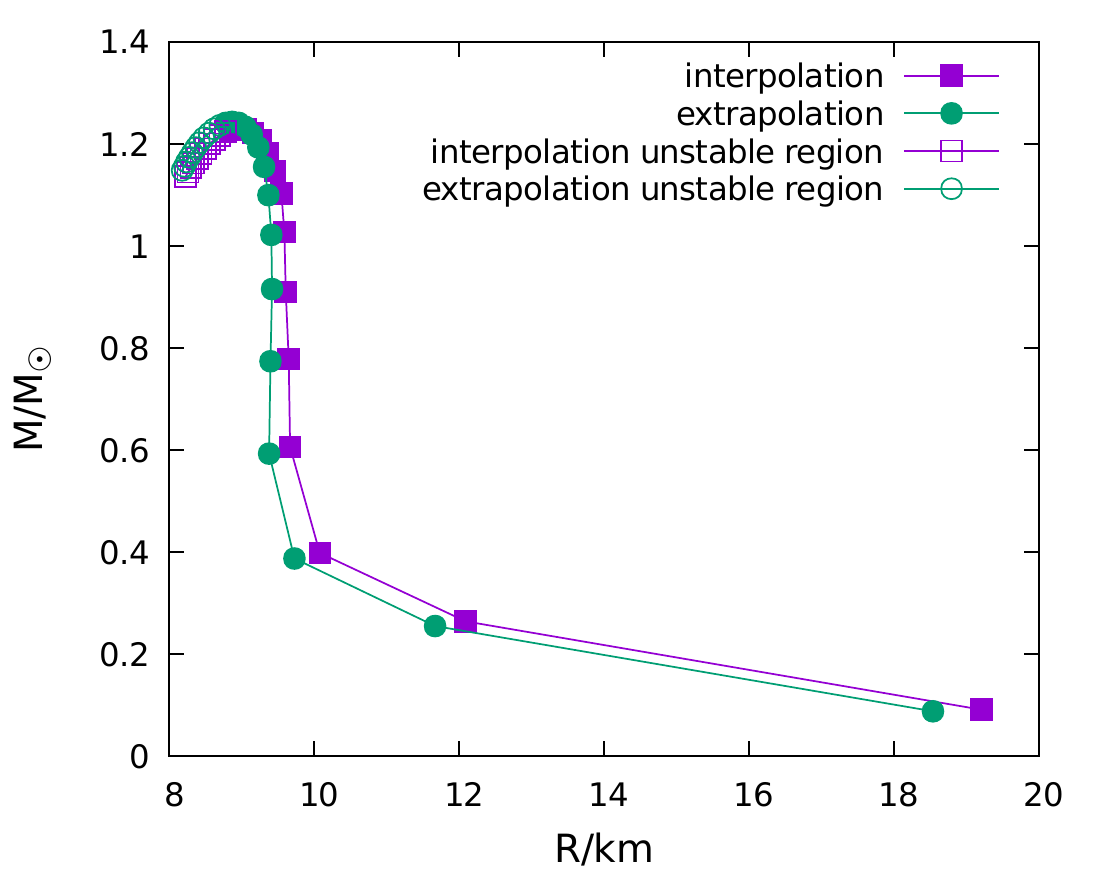}\\
\includegraphics[width=0.5\textwidth,type=pdf,ext=.pdf,read=.pdf]{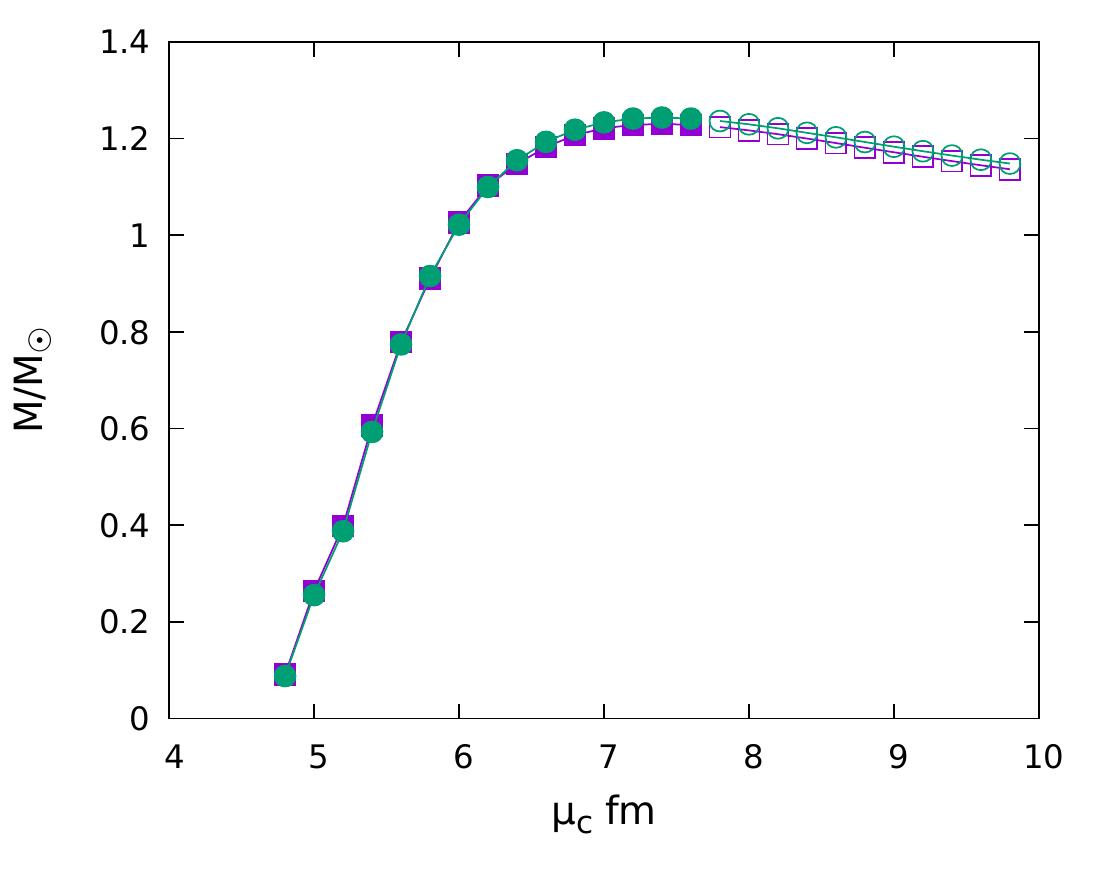}
\caption{\label{fig:n-mu2int} The red curve of the top-left panel shows the fit function passing through data points that differs from the interpolation by the dashed blue line. The top-right panel shows the implications for the mass-radius relation. The bottom panel gives the impact on the mass-central-chemical potential relation. All results for the light ensemble. Lines connecting the points are to guide the eye.} 
\end{figure}

In this context another question is how we perform the interpolation between the data points, especially as they in principle, due to statistical fluctuations, are not a smooth function. This is thus a test of the influence of the finite statistics and the discreteness of lattice results. The effect of an alternative fit, which uses more points at high density and which passes well thorough the lower density points, is shown in figure \ref{fig:n-mu2int}, together with its implications for the mass-radius relation. There is hardly any difference. This shows that an effective description of the equation of state using heavy fermions and modified parameters reproduces the full dynamics rather well. Thus, the replacement of the interpolation by the fit has much less systematic impact than the choice of the transition chemical potential $\mu_t$.

\subsection{Effects of uncertainties of the mass of the neutron and additional contributions to the energy density}

Another issue is that the mass of the neutron is actually determined within a certain statistical error in the lattice simulations \cite{Wellegehausen:2013cya}. This is here overshadowed by the fact the neutron's mass is used for the scale setting. Thus, the statistical error is actually an error on the lattice spacing used to give also the chemical potential physical units. Studying the impact of the statistical error on any of these quantities would, in dimensionless units, always give the same result. We therefore try to quantify it by varying the mass of the neutron in physical units within the corresponding error band, which is $\pm5$\% for the heavy ensemble and for the light ensemble $\pm8$\%,  slightly larger \cite{Wellegehausen:2013cya}. Note that a similar error would be induced if our assumption for the energy density to be given essentially by the vacuum rest mass of the neutron would be  incorrect at the same level. Thus, by varying the neutron mass, we actually study both kinds of influences, one from statistics, the other one from systematics.

\begin{figure}
\includegraphics[width=0.5\textwidth,type=pdf,ext=.pdf,read=.pdf]{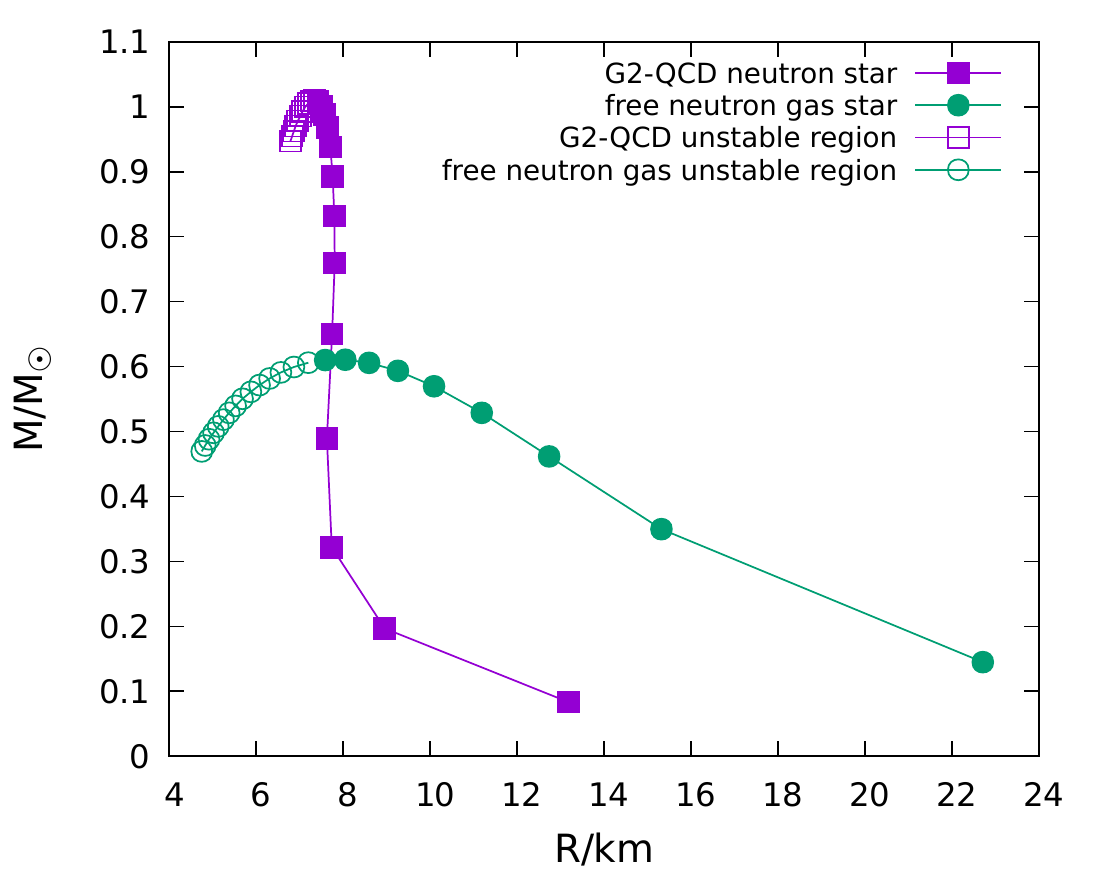}\includegraphics[width=0.5\textwidth,type=pdf,ext=.pdf,read=.pdf]{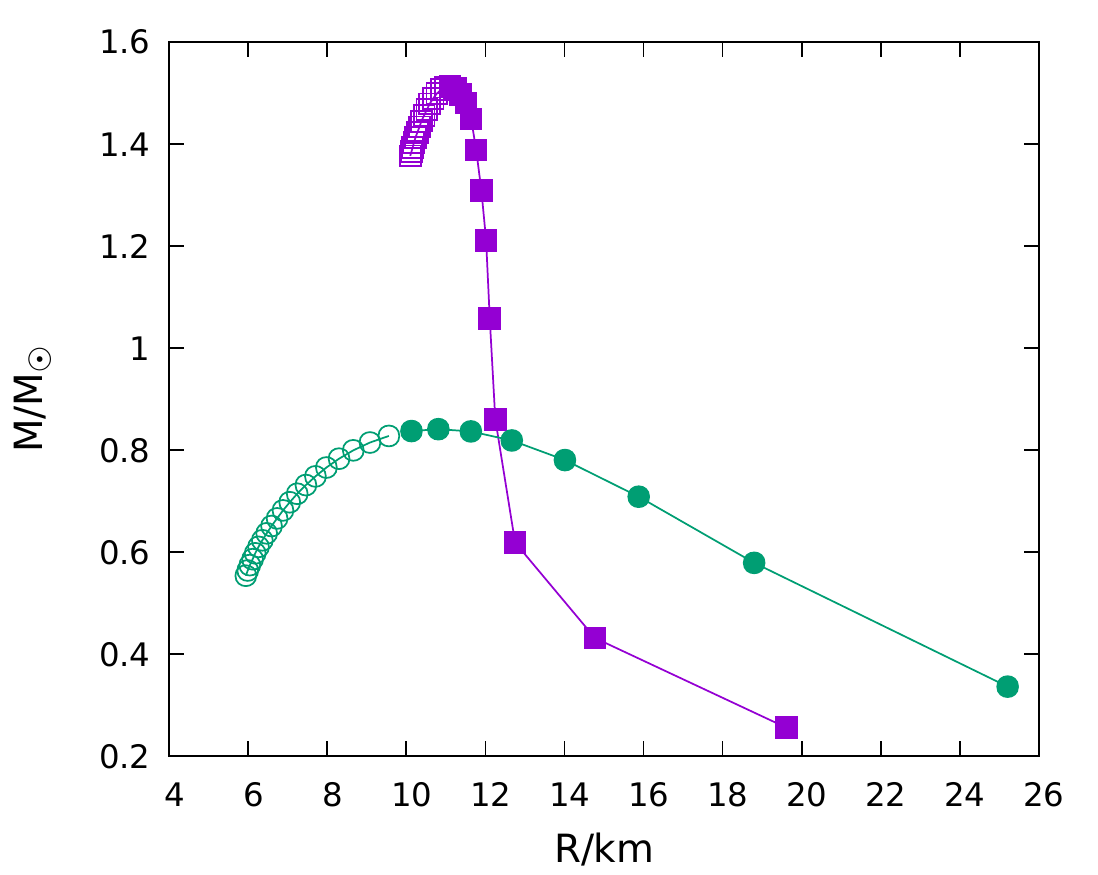}
\caption{\label{fig:mrmndep}Impact on the mass-radius relation of varying the neutron mass by $+8$\% (left panel) and $-8$\% (right panel) for the light ensemble (purple points) and for the free gas (green points). Note that the solar mass remained unchanged.} 
\end{figure}

The impact of this variation on the mass-radius relation is shown in figure \ref{fig:mrmndep}. The maximum mass decreases (increases) for a heavier (lighter) neutron. The latter is strongly influenced by the fact that another lattice data point in the baryon density becomes actually available when lowering the neutron mass by this amount, and thus a different low-energy fitting becomes necessary. This implies that details of the energy density and hadron masses can have a substantial quantitative influence, but the qualitative influence seems to be still not too extreme to invalidate the qualitative conclusions drawn in the main text.

\subsection{Effects of the assumptions about the equation of state}

\begin{figure}
\includegraphics[width=0.5\textwidth,type=pdf,ext=.pdf,read=.pdf]{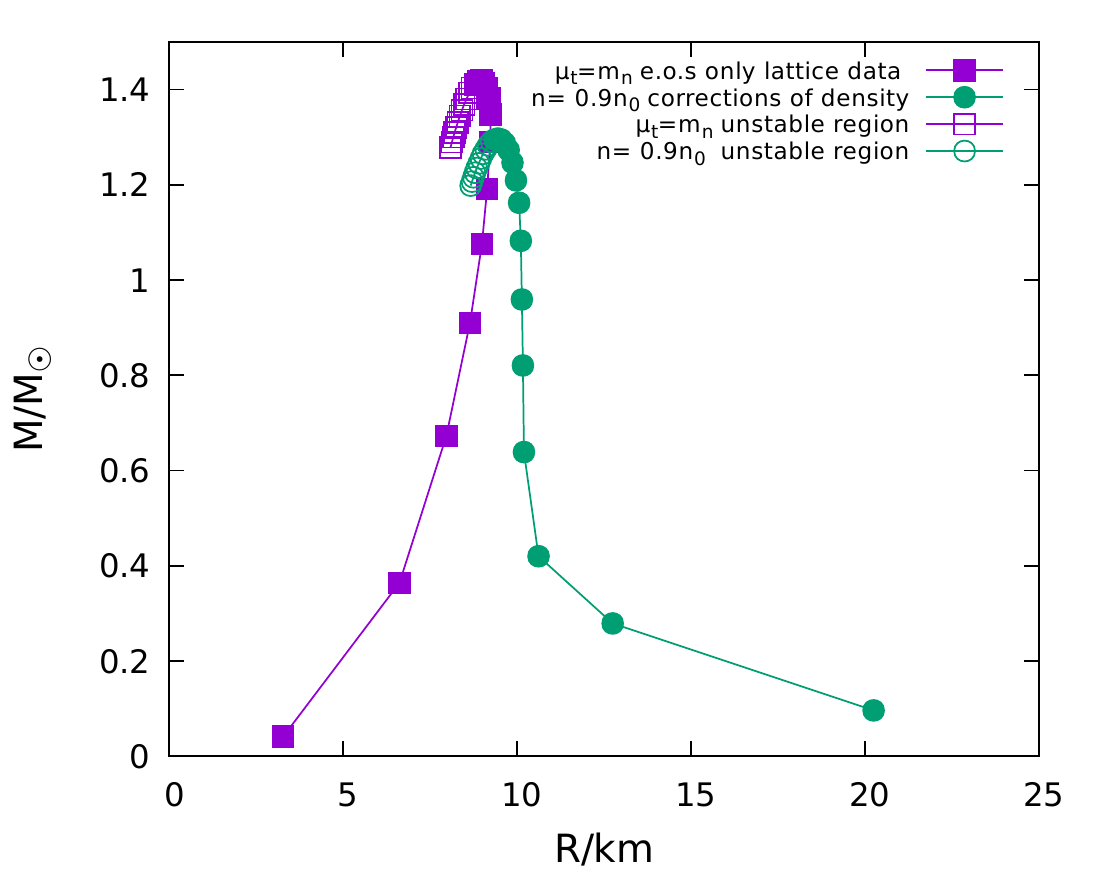}\includegraphics[width=0.5\textwidth,type=pdf,ext=.pdf,read=.pdf]{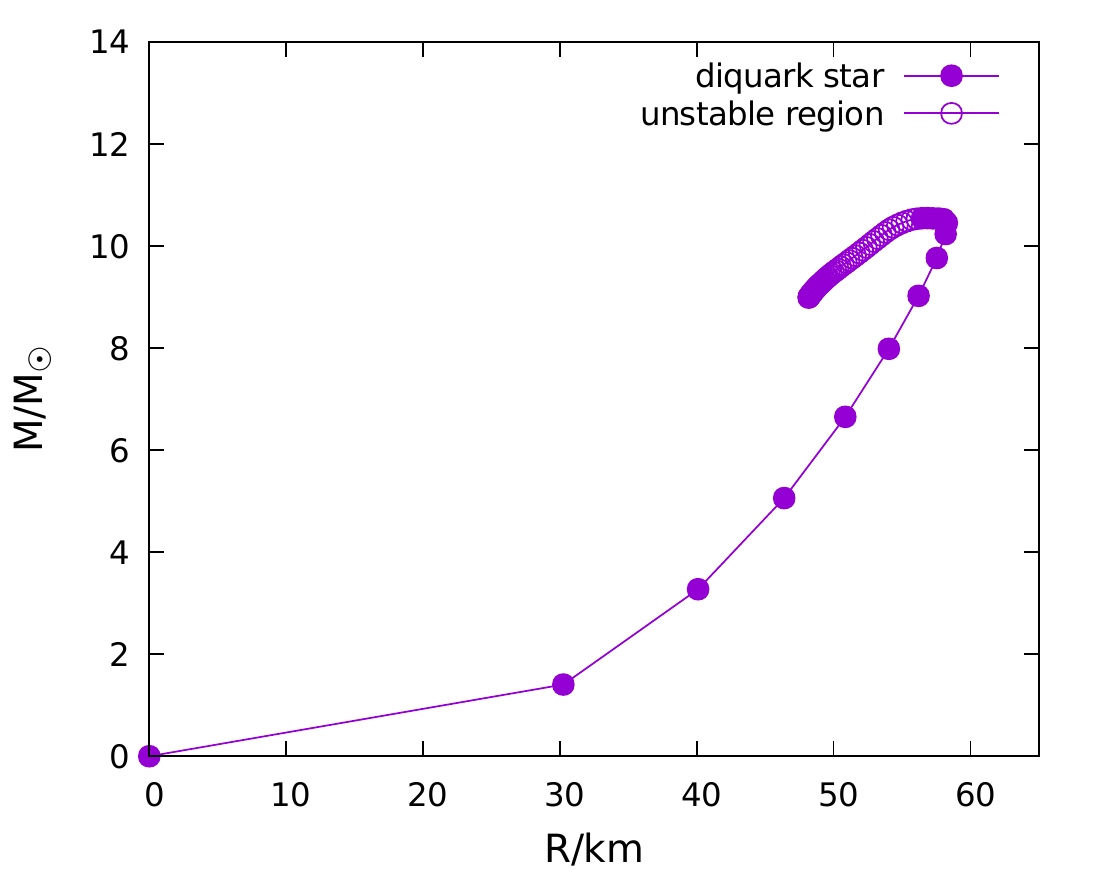}
\caption{\label{fig:lowd}Effects of altering the assumptions about the equation of state. The left-hand plot shows the results of assuming that all of the equation of state down to the crust is created from the neutron, allowing for additional interaction contributions, and for the case that the neutrons only contribute part of the baryon density. The right-hand plot shows the result if the full equation of state would be dominated by the diquarks. Note that the normalization of the mass axis is still in units of the ordinary solar mass.} 
\end{figure}

Arguably the most drastic approximation in our investigation is made by assuming that the baryon density is given entirely by the neutrons and goes into a free-gas behavior at low densities. Here, we relax these assumptions in three ways, with the results being shown in figure \ref{fig:lowd}:
\begin{itemize}
 \item Assuming that the actual baryon density is not entirely due to neutrons, but partly due to other baryons, which however do not contribute in the neutron star, is explored by rescaling $n\to xn$, with some number $x$. The results for $x=0.9$ in figure \ref{fig:lowd}, left panel, show no qualitative change.
 \item Assuming that the system is not of free-gas type close to the surface, but using instead the full baryon density down to the neutron mass is also shown in figure \ref{fig:lowd}, left panel. The energy density was still constructed in the same way. This changed little for the maximum mass and radius, but creates a 'wrong' bending at lower masses.
 \item Assuming the star to be made entirely out of diquarks, and thus using the baryon density down to the silver-blaze point, yields the results shown in figure \ref{fig:lowd}, right panel. We still assumed that the energy density is given by the diquark mass times the baryon density. This scenario can only be realized if the diquarks would form a Fermi surface. It is seen that this gives the same 'wrong' bending, and rescales the results roughly by the ratio of diquark mass to neutron mass. The latter is expected as then also normal stars would be made out of diquarks rather than nucleons, which we did not include.
\end{itemize}
In total, this always shows the same pattern: The shape close to the maximum mass is quite similar, and its value, up to rescaling effects, remains the same. Including a larger density at low chemical potential yields the change of bending. It is therefore mandatory to understand on the lattice what is the composition and the physics in this low-density regime, and especially to determine the energy density independently.

\section{Conclusion and outlook}\label{s:sum}

In this work, we have studied for the first time the structure of an idealized neutron star based only on lattice data of the model gauge theory G$_2$-QCD. Therefore we included the gauge degrees of freedom explicitly. But, due to the sign problem, it was necessary to use the model theory G$_2$-QCD rather than real QCD. Still, though the theory in question, G$_2$-QCD, is not ordinary QCD, it shares a multitude of features with QCD \cite{Maas:2012ts}. Besides the insight into the technical procedure, there may therefore still be hope that it is possible that some features of our results could carry over to real QCD.

Concerning the two questions we have raised in the introduction, we can answer them in the following way.

Concerning the second question, the results show a number of interesting features. In particular, the mass-radius relations becomes steeper and the maximum mass increases compared to the free-neutron gas case. The final mass is not too large, roughly 1.2 solar masses, with a systematic error margin of at least 0.1 solar masses. However, given the many approximations and that only a single flavor is present makes this rather small value not unexpected.

Taking it at face value, this could hint that either the masses of the quarks or the number of degrees of freedom play a significant role for the maximum neutron star mass. But on the other hand, the quark mass, measured by the Goldstone mass, seems to have no big influence on the results in the present case, at least for the two, not too small, Goldstone masses used here.

Less speculative and more visible in the data is that the phase change from a free gas at the surface to the interacting lattice data impresses itself into the mass-radius relation of the neutron star. It also shows signals in the internal structure of the neutron star, suggesting a layered structure for heavier neutron stars. Especially, there is an inner core containing most of the mass and the radius, and a smaller outer shell without strong interactions. This implies that already observing the mass-radius relation in sufficient detail could resolve the old question of whether a neutron star is a, more or less, monolithic object, or whether it contains a multitude of phases. If this mass-radius relation could be astronomically better measured, this would be an interesting test of these qualitative results.

These statements are, of course, only true to the extent as we assume that our results carry over to the exact case and that G$_2$-QCD is as similar to ordinary QCD as we expect from its microscopic similarity.

The insights gained for the first, at the current stage more important, question, can be summarized as a to-do list, in order of importance, for the lattice simulations:
\begin{itemize}
 \item The most important requirement is to obtain the energy density explicitly.
 \item The number of measured chemical potential values should be quite dense in the region relevant for the outer crust of the neutron star.
 \item It is necessary to determine when and how a Fermi surface forms, i.\ e.\ at which chemical potential can a star start to withstand the gravitational pressure. This requires to know the relevant degrees of freedom at this point.
 \item What is the relevance of bosonic hadrons? What are the various transitions between fermionic and bosonic states? How do they contribute to the number and energy densities?
 \item Do the properties of the particles change with density?
 \item The generic requirements for lattice calculations of larger and finer lattices and better statistics hold, as always.
 \item In the long run, the calculation of bulk properties to verify the viability of the ideal fluid approximation would be helpful, as would be any insight on possible (color)superconducting phases \cite{Buballa:2003qv}.
\end{itemize}
Note that even if QCD does not have diquarks, similar questions will arise as well. E.\ g.\ the question of the fraction of pions or excited nucleons inside a moderately dense regime will be important.

This was a first, exploratory investigation. Also the lattice data have been more exploratory than quantitative \cite{Wellegehausen:2013cya}. Thus, there are many avenues to improve the results presented here. A major step would be to resolve the to-do list. However, detailed studies show that this will be a very demanding task, even without the sign problem \cite{Wellegehausen:2015iea,Wellegehausen:2017gba}, not to mention QCD itself.

An alternative would be to augment the lattice results with results from other sources to obtain a more realistic equation of state. At the current time, no such input is available for G$_2$-QCD. However, based on our experiences here it is likely that this will be necessary also for a QCD calculation in the time to come, even if the sign problem would be fully resolved. Thus, it may be worthwhile to develop corresponding tools also for G$_2$-QCD to start studying the interplay of lattice with non-lattice methods to build neutron stars.

Following such an avenue would help to prepare for understanding results of gravitational wave signals of neutron star mergers. After all, having a reliable equation of state of G$_2$-QCD would allow to even simulate G$_2$-QCD neutron star mergers, and thus understand how the inner structure of neutron stars imprints on the gravitational wave signal, as it is expected to do in QCD  \cite{DelPozzo:2013ala}. This is, of course, a highly demanding task, and most probably will not be achieved in the near future, given the available resources.\\

\no{\bf Acknowledgements}

We acknowledge helpful discussions on G$_2$-QCD with B.\ Wellegehausen, L.\ von Smekal, and A.\ Wipf. O.\ H.\ was supported by the FWF doctoral school W1203-N16.

\bibliographystyle{bibstyle}
\bibliography{bib}

\begin{thebibliography}{10}

\bibitem{Glendenning:1997wn}
N.~K. Glendenning,
\newblock {\em {Compact stars: Nuclear physics, particle physics, and general
  relativity}} (, 1997).

\bibitem{Steiner:2010fz}
A.~W. Steiner, J.~M. Lattimer, and E.~F. Brown,
\newblock Astrophys. J. {\bf 722}, 33 (2010), 1005.0811.

\bibitem{Kapusta:2006pm}
J.~I. Kapusta and C.~Gale,
\newblock {\em {Finite-temperature field theory: Principles and applications}}
  (Cambridge University Press, Cambridge, 2006).

\bibitem{Lattimer:2006xb}
J.~M. Lattimer and M.~Prakash,
\newblock Phys. Rept. {\bf 442}, 109 (2007), astro-ph/0612440.

\bibitem{Abbott:2016blz}
Virgo, LIGO Scientific, B.~P. Abbott {\em et~al.},
\newblock Phys. Rev. Lett. {\bf 116}, 061102 (2016), 1602.03837.

\bibitem{DelPozzo:2013ala}
W.~Del~Pozzo, T.~G.~F. Li, M.~Agathos, C.~Van Den~Broeck, and S.~Vitale,
\newblock Phys. Rev. Lett. {\bf 111}, 071101 (2013), 1307.8338.

\bibitem{Friman:2011zz}
B.~Friman {\em et~al.},
\newblock Lect.Notes Phys. {\bf 814}, 1 (2011).

\bibitem{Gattringer:2010zz}
C.~Gattringer and C.~B. Lang,
\newblock {\em Quantum chromodynamics on the lattice} (Lect. Notes Phys.,
  2010).

\bibitem{deForcrand:2010ys}
P.~de~Forcrand,
\newblock PoS {\bf LAT2009}, 010 (2009), 1005.0539.

\bibitem{Aarts:2017vrv}
G.~Aarts, E.~Seiler, D.~Sexty, and I.-O. Stamatescu,
\newblock (2017), 1701.02322.

\bibitem{Leupold:2011zz}
S.~Leupold {\em et~al.},
\newblock Lect.Notes Phys. {\bf 814}, 39 (2011).

\bibitem{Buballa:2003qv}
M.~Buballa,
\newblock Phys.Rept. {\bf 407}, 205 (2005), hep-ph/0402234.

\bibitem{Pawlowski:2010ht}
J.~M. Pawlowski,
\newblock AIP Conf.Proc. {\bf 1343}, 75 (2010), 1012.5075.

\bibitem{Braun:2011pp}
J.~Braun,
\newblock J.Phys.G {\bf G39}, 033001 (2012), 1108.4449.

\bibitem{Holland:2003jy}
K.~Holland, P.~Minkowski, M.~Pepe, and U.~J. Wiese,
\newblock Nucl. Phys. {\bf B668}, 207 (2003), hep-lat/0302023.

\bibitem{Maas:2012wr}
A.~Maas, L.~von Smekal, B.~Wellegehausen, and A.~Wipf,
\newblock Phys.Rev. {\bf D86}, 111901 (2012), 1203.5653.

\bibitem{Maas:2012ts}
A.~Maas and B.~H. Wellegehausen,
\newblock PoS {\bf LATTICE2012}, 080 (2012), 1210.7950.

\bibitem{Wellegehausen:2013cya}
B.~H. Wellegehausen, A.~Maas, A.~Wipf, and L.~von Smekal,
\newblock Phys.Rev. {\bf D89}, 056007 (2014), 1312.5579.

\bibitem{Wellegehausen:2015iea}
B.~H. Wellegehausen and L.~von Smekal,
\newblock PoS {\bf LATTICE2014}, 177 (2015), 1501.06706.

\bibitem{Pepe:2006er}
M.~Pepe and U.~J. Wiese,
\newblock Nucl. Phys. {\bf B768}, 21 (2007), hep-lat/0610076.

\bibitem{Greensite:2006sm}
J.~Greensite, K.~Langfeld, {\v S}.~Olejn\'ik, H.~Reinhardt, and T.~Tok,
\newblock Phys. Rev. {\bf D75}, 034501 (2007), hep-lat/0609050.

\bibitem{Maas:2007af}
A.~Maas and {\v S}.~Olejn\'ik,
\newblock JHEP {\bf 02}, 070 (2008), 0711.1451.

\bibitem{Cossu:2007dk}
G.~Cossu, M.~D'Elia, A.~Di~Giacomo, B.~Lucini, and C.~Pica,
\newblock JHEP {\bf 10}, 100 (2007), 0709.0669.

\bibitem{Danzer:2008bk}
J.~Danzer, C.~Gattringer, and A.~Maas,
\newblock JHEP {\bf 01}, 024 (2009), 0810.3973.

\bibitem{Liptak:2008gx}
L.~Liptak and {\v S}.~Olejn\'ik,
\newblock Phys. Rev. {\bf D78}, 074501 (2008), 0807.1390.

\bibitem{Maas:2010qw}
A.~Maas,
\newblock JHEP {\bf 02}, 076 (2011), 1012.4284.

\bibitem{Wellegehausen:2009rq}
B.~H. Wellegehausen, A.~Wipf, and C.~Wozar,
\newblock Phys. Rev. {\bf D80}, 065028 (2009), 0907.1450.

\bibitem{Wellegehausen:2010ai}
B.~H. Wellegehausen, A.~Wipf, and C.~Wozar,
\newblock Phys.Rev. {\bf D83}, 016001 (2011), 1006.2305.

\bibitem{Ilgenfritz:2012aa}
E.-M. Ilgenfritz and A.~Maas,
\newblock Phys.Rev. {\bf D86}, 114508 (2012), 1210.5963.

\bibitem{Bruno:2014rxa}
M.~Bruno, M.~Caselle, M.~Panero, and R.~Pellegrini,
\newblock JHEP {\bf 03}, 057 (2015), 1409.8305.

\bibitem{Bonati:2015uga}
C.~Bonati,
\newblock JHEP {\bf 03}, 006 (2015), 1501.01172.

\bibitem{Wellegehausen:2011jz}
B.~H. Wellegehausen,
\newblock PoS {\bf LATTICE2011}, 266 (2011), 1111.0496.

\bibitem{Hajizadeh:2016jvj}
O.~Hajizadeh and A.~Maas,
\newblock PoS {\bf LATTICE2016}, 358 (2016), 1609.06979.

\bibitem{Kogut:2000ek}
J.~Kogut, M.~A. Stephanov, D.~Toublan, J.~Verbaarschot, and A.~Zhitnitsky,
\newblock Nucl.Phys. {\bf B582}, 477 (2000), hep-ph/0001171.

\bibitem{vonSmekal:2012vx}
L.~von Smekal,
\newblock Nucl. Phys. Proc. Suppl. {\bf 228}, 179 (2012), 1205.4205.

\bibitem{Kitazawa:2016dsl}
M.~Kitazawa, T.~Iritani, M.~Asakawa, T.~Hatsuda, and H.~Suzuki,
\newblock Phys. Rev. {\bf D94}, 114512 (2016), 1610.07810.

\bibitem{Gandolfi:2013baa}
S.~Gandolfi, J.~Carlson, S.~Reddy, A.~W. Steiner, and R.~B. Wiringa,
\newblock Eur. Phys. J. {\bf A50}, 10 (2014), 1307.5815.

\bibitem{Cohen:2003kd}
T.~D.~. Cohen,
\newblock Phys.Rev.Lett. {\bf 91}, 222001 (2003), hep-ph/0307089.

\bibitem{Wellegehausen:2017gba}
B.~H. Wellegehausen and L.~von Smekal,
\newblock PoS {\bf LATTICE2016}, 078 (2016), 1702.00238.

\end{thebibliography}

\end{document}